\documentclass[aps,prd,twocolumn,showpacs,groupedaddress,nofootinbib]{revtex4}
\usepackage{graphicx}  % needed for figures
\usepackage{dcolumn}   % needed for some tables
\usepackage{bm}        % for math
\usepackage[english]{babel}
\usepackage{amsfonts,amsmath,amssymb,mathrsfs}
\usepackage{epstopdf}
\usepackage{subfigure}
\usepackage{color}

\def\a{\alpha}
\def\r{\rho}
\def\s{\sigma}
\def\t{\tau}
\def\m{\mu}
\def\n{\nu}
\def\k{\kappa}
\def\th{\theta}
\def\g{\gamma}\def\G{\Gamma}
\def\L{\Lambda}\def\l{\lambda}
\def\D{\Delta}
\def\la{\langle}
\def\ra{\rangle}
\def\o{\omega}\def\O{\Omega}
\def\d{\delta}
\def\p{\partial}

\def\half{\textstyle{\frac{1}{2}}}

\def\bdoc{\begin{document}}
\def\edoc{\end{document}}

\def\beq{\begin{equation}}
\def\eeq{\end{equation}}
\def\bea{\begin{eqnarray}}
\def\eea{\end{eqnarray}}
\def\ben{\begin{enumerate}}
\def\een{\end{enumerate}}
\def\la{\langle}
\def\ra{\rangle}
\def\a{\alpha}
\def\b{\beta}
\def\g{\gamma}\def\G{\Gamma}
\def\d{\delta}\def\D{\Delta}
\def\e{\epsilon}
\def\z{\zeta}

\def\th{\theta}
\def\k{\kappa}
\def\l{\lambda}
\def\m{\mu}
\def\n{\nu}
\def\o{\omega}
\def\p{\pi}
\def\r{\rho}
\def\s{\sigma}
\def\t{\tau}
\def\L{{\cal L}}
\def\S{\Sigma }
\def\gsim{\; \raisebox{-.8ex}{$\stackrel{\textstyle >}{\sim}$}\;}
\def\lsim{\; \raisebox{-.8ex}{$\stackrel{\textstyle <}{\sim}$}\;}
\def\gtrsim{\gsim}
\def\lessim{\lsim}
\def\loc{{\rm local}}
\def\vm{v_{\rm max}}
\def\bh{\bar{h}}
\def\del{\partial}
\def\nab{\nabla}
\def\half{{\textstyle{\frac{1}{2}}}}
\def\fourth{{\textstyle{\frac{1}{4}}}}

\def\bD{{\bf D}}
\def\bE{{\bf E}}
\def\bF{{\bf F}}
\def\bB{{\bf B}}
\def\bP{{\bf P}}
\def\bV{{\bf v}}
\def\bv{{\bf v}}
\def\bx{{\bf x}}
\def\by{{\bf y}}
\def\bz{{\bf z}}
\def\ba{{\bf a}}
\def\bd{{\bf d}}
\def\bs{{\bf s}}
\def\bn{{\bf n}}
\def\bp{{\bf p}}

\def\O{\Omega}

\def\br{{\bf r}}
\def\bnab{{\bf \nab}}

\def\tE{\tilde{E}}
\def\tL{\tilde{L}}
\def\Horava{Ho\v{r}ava }

\begin{document}

\title{Black holes in Einstein-aether and Ho\v rava--Lifshitz gravity}
\author{Enrico Barausse$^1$, Ted Jacobson$^1$ and Thomas P. Sotiriou$^2$}
\affiliation{${}^1$Center for Fundamental Physics,  University of Maryland, College Park, MD 20742-4111, USA\\ ${}^2$Department of Applied Mathematics and Theoretical Physics, Centre for Mathematical Sciences, University of Cambridge, Wilberforce Road, CB3 0WA, Cambridge, UK}
\date{\today} 
\begin{abstract}
We study spherical black-hole solutions 
in Einstein-aether theory, a Lorentz-violating gravitational theory consisting
of General Relativity with a dynamical unit timelike vector (the ``aether'') that defines a preferred
timelike direction. These are also solutions to the infrared limit of Ho\v rava--Lifshitz gravity.
We explore parameter values of the two theories  where all 
presently known experimental constraints are satisfied,
and find that spherical black-hole solutions of the type expected to form 
by gravitational collapse exist for all those parameters. 
Outside the metric horizon, the deviations  away from the Schwarzschild metric 
are typically no more than a few percent
for most of the explored parameter regions,
which makes them difficult to observe with electromagnetic probes, 
but in principle within reach of future gravitational-wave detectors. 
Remarkably, we find that the solutions possess  
a universal horizon, not far inside the metric horizon, 
that traps waves of any speed relative to the aether.
A notion of black hole thus persists in these theories, even in 
the presence of arbitrarily high propagation speeds. 
\end{abstract}  
\maketitle

\section{Introduction}

Lorentz invariance is believed to be a fundamental symmetry of physical theories. Indeed, there are severe observational constraints on Lorentz-violating effects in the matter sector \cite{Mattingly:2005re,Liberati:2009pf}. On the other hand, in the much more weakly coupled 
gravitational sector such constraints are generically far weaker. Therefore, 
it is interesting to test Lorentz symmetry further in gravitational phenomena.

To do that in a well-defined way one must
consider some candidate 
Lorentz-violating (LV) gravitational theory as a low energy effective theory. To violate Lorentz symmetry 
and still be manifestly diffeomorphism invariant, such a theory should include, apart from the metric, 
some dynamical field that can define a preferred frame at the level of the solution. A unit timelike vector 
field is 
an example which breaks local boost but not local rotation symmetries.
The most general theory one can construct by 
coupling this field to general relativity (GR) at second order in derivatives is called Einstein-aether 
theory (\ae-theory) \cite{Gasperini, Jacobson:2000xp, Jacobson:2008aj}.
The vector field is referred to as the aether.

Apart from providing a 
test bed for constraining Lorentz violations in the gravitational sector, \ae-theory 
is
an interesting theoretical laboratory to explore preferred frame effects without having to give up diffeomorphism invariance. Seen as a low energy effective field theory \cite{Withers:2009qg},
 it can be thought of as encapsulating LV effects 
that might arise in a more fundamental quantum gravity theory. Its viability as a low energy effective theory of gravity has be extensively tested against various different observations, and up to date there seems to be a significant portion of the parameter space for which the predictions agree with all current experimental evidence (see Ref.\ \cite{Jacobson:2008aj} for a review).

Recently, another proposal for a LV gravity theory has received a lot of attention, Ho\v{r}ava--Lifshitz (HL) gravity \cite{Horava:2009uw}. This is not supposed to be just an effective field theory. 
There is hope that it can constitute instead an actual UV completion of general relativity, as it appears to be power-counting renormalizable. This is achieved by adding higher order spatial derivatives,  
without adding higher order time derivatives, which lead to a suitable modification of the propagator. 
In HL gravity this involves
the existence of a preferred 
spacelike foliation of spacetime,
which is described by a scalar field. 

The dynamics of this scalar (or the lack thereof) 
can lead to various problems in restricted versions of the theory 
such as instabilities, over-constrained evolution, 
and strong coupling at low energies \cite{bunch,Sotiriou:2009bx}. 
However, as pointed in Ref.\ \cite{Blas:2009qj},
 the dynamical behavior of the scalar is drastically improved when 
 one consistently includes in the action 
all the possible operators that are allowed by the symmetry of the theory. 
There is only one of these operators at low energies, and its 
presence suffices to ensure dynamical consistency and to push strong coupling to sufficiently high energies, so that the theory makes sense as an effective theory. 
Strong coupling at high energies, which would still be a threat for UV completeness, persists in general~\cite{Papazoglou:2009fj,Kimpton:2010xi},  but it appears to be avoided by assigning a specific hierarchy to the scales suppressing the lower and higher 
order operators \cite{Blas:2009ck}. 
There are many things to be worked out before HL gravity can be considered a viable UV complete gravity theory, such as renormalizability beyond power-counting, renormalization group flow of the various running couplings (so far infrared (IR) viability hinges on the hope that several parameters will run 
or can be tuned
to desired values), various phenomenological aspects and constraints, etc. However, to date it certainly constitutes an interesting candidate.

Given that \ae-theory is a quite generic effective theory of LV gravity with a single preferred
local timelike direction, 
it is reasonable to expect that the low energy limit of HL gravity will bear some resemblance to it. Indeed, it was remarked in Ref.\ \cite{Blas:2009ck} (see also Ref.\ \cite{Germani:2009yt}) and then 
fully demonstrated 
in Ref.~\cite{Jacobson:2010mx} that, in the limit where higher than second order operators can be neglected, HL gravity is 
equivalent to 
\ae-theory with the extra condition
that the aether is hypersurface orthogonal at the level of the action. 
Therefore, the IR limit of HL gravity can be understood as a limiting version of a well studied theory (some, though not all results carry over), which also gives hope that HL can be viable from the IR perspective. On the other hand, \ae-theory, or at least some version of it, acquires some more robust theoretical motivation as a low energy limit of a 
potentially 
UV complete gravity theory.

One of the results that carry over between the two theories is  
spherically symmetric solutions. This is because all spherically symmetric aether fields are hypersurface orthogonal and, hence, all spherically 
symmetric solutions of \ae-theory will also be solutions of the IR limit of HL gravity \cite{Jacobson:2010mx}. The converse holds for solutions with a
regular center~\cite{Blas:2010hb}, but without this condition there may be additional
HL solutions. Here we do not address that possibility, and instead focus entirely on
black hole solutions to \ae-theory. Such solutions 
are
interesting mathematically as black holes 
``dressed'' by 
the aether. They can also be used in comparing 
these theories
with observations of astrophysical black holes
(though only as a first step since 
our solutions are non-rotating).
%they do not allow for rotation). 

Motivated by this last point, 
our intention is to find, for each set of coupling parameters that meet
current observational constraints, 
the unique static, spherically symmetric, vacuum, asymptotically flat 
black hole solution of \ae-theory that forms from collapse.
Since different modes travel with different speeds in both \ae-theory and 
HL gravity, the 
definition of a ``black hole'' in these theories is potentially ambiguous. 
At the outset, our working definition will be that a black hole possesses both
a metric horizon and a spin-0 mode horizon. We find, however, that 
in all of the several cases we have
checked
these solutions actually possess a ``universal horizon,'' 
not far inside the metric horizon,
that traps modes of any speed.

We determine the black hole solutions numerically (using Mathematica),
because obtaining analytic solutions of the equations does not appear to be feasible. 
As will be explained below, 
the restriction to black holes that form from collapse amounts to the
requirement that the horizon for the superluminal spin-0 mode, which lies
inside the metric horizon, is nonsingular. 
Our analysis generalizes the results of Ref.~\cite{Eling:2006ec} 
which focused on a restricted, non-viable, choice of coupling parameters in the action,
and the results of Ref.~\cite{Tamaki:2007kz}, which considered 
observationally viable coupling parameters, but did not impose 
regularity of the spin-0 horizon. Though other work has been done on
black hole solutions in HL gravity, it was not in the same version of the theory considered here, 
but in versions which either impose projectability (the lapse function is forced to be 
space-independent), {\em e.g.}~Ref.~\cite{Greenwald:2009kp},
 or consider only a reduced set of terms in the action, {\em e.g.}~Refs.~\cite{Lu:2009em,Kehagias:2009is,Kiritsis:2009rx}.\footnote{The solutions found in Ref.~\cite{Kiritsis:2009vz} for HL gravity had been found earlier in Ref.~\cite{Eling:2006df} as solutions to \ae-theory and are not actually black hole solutions.}
 In contrast, as explained above, we include the complete set of terms in the IR limit.
This changes the nature of the black hole solutions.

The rest of the paper is organized as follows. In section \ref{background} we briefly review \ae-theory and HL gravity, as well as their relation. In section \ref{spherical} we discuss the characteristics of generic spherically symmetric, asymptotically flat, black hole solutions and explain how the problem of generating numerical solutions can be set up. In section \ref{parameterspace} we summarize the various constraints we will take into account in order to restrict the parameter space in both \ae-theory and HL-gravity. We present and discuss numerical results in section \ref{results}. Section \ref{concs} contains our conclusions. 

In this paper, we denote spacetime indices by Greek letters, and spatial indices by Latin letters. We also adopt the spacetime signature 
$({+}{-}{-}{-})$
and set $c=1$.

\section{\AE-theory and HL gravity: brief overview}
\label{background}

The most general action for \ae-theory, up to total derivative terms and setting aside matter couplings, is
\beq \label{S}
S_{\ae} = \frac{1}{16\pi G_{\ae}}\int \sqrt{-g}~ (-R + L_{\ae})
~d^{4}x \eeq
where $R$ is the 4D Ricci scalar of the metric $g_{ab}$, $g$ is the determinant of the metric and
\beq \label{Lae}
L_{\ae} = -M^{\a\b}{}_{\m\n} \nabla_\a u^\m \nabla_\b u^\n,
\eeq
with $M^{\a\b}{}_{\m\n}$ defined as
\beq M^{\a\b}{}_{\m\n} = c_1 g^{\a\b}g_{\m\n}+c_2\d^{\a}_{\m}\d^{\b}_{\n}
+c_3 \d^{\a}_{\n}\d^{\b}_{\m}+c_4 u^\a u^\b g_{\m\n}. 
\eeq
The $c_i$ are dimensionless coupling constants,
and it is assumed that $u^\m$ is constrained to be a unit timelike
vector, $g_{\m\n}u^\m u^\n=1$. This constraint can be explicitly imposed 
with a Lagrange multiplier term $\l(g_{\m\n}u^\m u^\n-1)$ in the action.
Note that
since the covariant derivative $\nabla_\a u^\m$ involves
derivatives of the metric through the connection components, and
since the unit vector is nowhere vanishing, the terms quadratic in
$\nabla u$ also modify the kinetic terms for the metric.
One consequence of this is that 
the constant $G_{\ae}$ is related to Newton's constant, 
as defined in the Newtonian
limit, by \cite{Carroll:2004ai}
\beq
G_N = \frac{G_{\ae}}{1-(c_1+c_4)/2}.
\eeq

Varying the action (\ref{S}) with respect to the metric yields
\beq\label{gfe}
G_{\a\b} = T^{\ae}_{\a\b}
\eeq
where $G_{\a\b}=R_{\a\b}-R g_{\a\b}/2$ 
is the usual Einstein tensor.  $T^{\ae}_{\a\b}$ denotes the
aether stress-energy tensor
\bea\label{Tae}
T^{\ae}_{\a\b}&=&\nabla_\m\left(J^{\phantom{(\a}\m}_{(\a}u_{\b)}-J^\m_{\phantom{\m}(\a}u_{\b)}-J_{(\a\b)}u^\m\right)\nonumber\\
&&+c_1\,\left[ (\nabla_\m u_\a)(\nabla^\m u_\b)-(\nabla_\a u_\m)(\nabla_\b u^\m) \right]\nonumber\\
&&+\left[ u_\n(\nabla_\m J^{\m\n})-c_4 \dot{u}^2 \right] u_\a u_\b\nonumber\\
&&+c_4 \dot{u}_\a \dot{u}_\b-\frac{1}{2} L_{\ae} g_{\a\b}\,,
\eea
where
\beq
J^\a_{\phantom{a}\m}=M^{\a\b}_{\phantom{ab}\m\n} \nabla_\b u^\m\,
\eeq
and $\dot{u}_\n=u^\m\nabla_\m u_\n$. Variation with respect to $u^\mu$ yields
\beq
\label{AEeq}
\left(\nabla_\a J^{\a\n}-c_4\dot{u}_\a\nabla^\n u^\a\right) \left(g_{\mu\nu}-u_\m u_\n\right)=0\,.
\eeq
The Lagrange multiplier $\l$ has been eliminated from these equations
by solving for it using the aether field equation.

Suppose now we want to impose the restriction that the aether be 
hypersurface orthogonal. Locally, this amounts to saying that there exists 
a function $T$ for which
\beq
\label{ho}
u_\a=\frac{\partial_\a T}{\sqrt{g^{\m\n}\partial_\m T \partial_\n T}}\,,
\eeq 
where we have taken into account the unit constraint on the aether.
If this form for the aether is substituted into the action 
(\ref{S}), then one obtains a new theory, with fewer degrees of freedom,
which is in fact identical to the IR limit of HL gravity. It appears
from (\ref{Lae}) and (\ref{ho}) that the resulting action would lead to equations
of motion with fourth-order derivatives, and non-polynomial dependence
on the derivatives.
However, we can choose $T$ itself to be the time coordinate 
$t$, in which case we have 
\beq\label{preferred foliation}
u_\a=\delta_{\a}^{T} (g^{TT})^{-1/2}=N\delta_{\a}^{T}  \,,
\eeq
where $N=(g^{TT})^{-1/2}$ is the lapse function.
Moreover, the equation of motion that would come from 
variation of $T$ is identically satisfied when the metric and 
matter equations of motion are imposed, hence 
this gauge choice can be made before varying $T$ 
 (see Ref.\ \cite{Jacobson:2010mx} for more details).

This hypersurface-orthogonal \ae-theory action
then takes the form
\beq\label{SBPSH}
S_{h.o.\ae}=\frac{1}{16\pi G_{H}}\!\int dT d^3x \, N\sqrt{h} \, L_2
\eeq
with
\beq\label{L2}
L_2
=K_{ij}K^{ij} - \lambda K^2 
+ \xi {}^{(3)}\!R + \eta a_ia^i,
\eeq
where  $K^{ij}$ is the extrinsic curvature 
of each 
constant $T$ surface,
 $h_{ij}$ is the induced 
 spatial
 metric, ${}^{(3)}\!R$ its Ricci curvature, and
\beq 
a_i=\partial_i \ln{N}\,
\eeq
is the spatial projection of the acceleration of 
the
normal congruence, {\it i.e.} the
acceleration of the aether flow.
The correspondence of the various parameters 
is\footnote{A rescaling of the spatial metric 
can be used to set $\xi=1$ in the absence of matter, as sometimes done 
in the literature. This is no longer possible once matter is present.}
\beq
\label{HLpar}
\frac{G_H}{G_{\ae}}=\xi=\frac{1}{1-c_{13}}, \quad \lambda=\frac{1+c_2}{1-c_{13}},\quad \eta=\frac{c_{14}}{1-c_{13}},
\eeq
where we use the notation $c_{ij}=c_i+c_j$.

Note that the coupling constants $c_1$, $c_3$ and $c_4$ enter only
through the combinations $c_{13}$ and $c_{14}$. This can be traced to 
a redundancy in the terms of the action when the aether is hypersurface
orthogonal. This is relevant both for HL gravity, and for \ae-theory in 
spherical symmetry, since any spherically symmetric vector field is hypersurface
orthogonal. The twist 
$\omega_\a=\epsilon_{\a\b\g\d} u^\b \nabla^\g u^\d$ vanishes
for any such vector field. For a unit vector field, the square of the twist 
 is given by
\bea
\label{twist}
&&\omega_\a\omega^\a=-(\nabla_\a u_\b)(\nabla^\a u^\b)+(\nabla_\a u_\b)(\nabla^\b u^\a)\nonumber\\&&\qquad\qquad\qquad+(u^\b \nabla_\b u_\a)(u^\m \nabla_\m u^\a).
\eea
As far as solutions with zero twist are concerned, any multiple of 
$\omega_\a\omega^\a$ can be added to the action (\ref{S}) without changing the
solutions. For example, adding $c_1 \omega_\a\omega^\a$ results in 
new couplings
$c_1'=0$, $c_3'=c_{13}$, and $c_4'=c_{14}$. This shows that 
the hypersurface orthogonal solutions depend only on $c_2$,
$c_{13}$, and $c_{14}$.
Alternatively,  one can subtract $c_4 \omega_\a\omega^\a$ from the action,
eliminating the $c_4$ term. We will do this in the calculations that follow.

Having fixed the $T$ coordinate up to a global reparametrization
(under which $u_\a$ (\ref{ho}) is invariant),
the symmetry of the theory is reduced to that of ``foliation preserving 
diffeomorphisms," i.e.\ space-independent
time reparametrization together with time-dependent spatial diffeomorphisms,
\beq
\label{fpd}
T\rightarrow T'(T), \quad x^i\rightarrow x'^i(x^i,T).
\eeq
Under these transformations $N' dT'= N dT$, and 
$a'_i = a_i$. The action (\ref{SBPSH}) is the most general 
one that is invariant under these symmetries and 
involves no more than two derivatives of $N$ and $g_{ij}$.

As already mentioned, the above theory is the IR limit of
HL gravity. The full HL action is of the form
\beq\label{SBPSHfull}
S_{HL}= \frac{1}{16\pi G_{H}}\int dT d^3x \, N\sqrt{h}(~L_2+\frac{1}{M_\star^2}L_4+\frac{1}{M_\star^4}L_6)\,,
\eeq
where 
$M_\star$ is a new mass scale, 
and $L_4$ and $L_6$ include all the 
foliation preserving diffeomorphism invariant scalar functions of $a_i$ and $h_{ij}$   
of 4th and 6th order in the spatial derivatives, 
respectively.
The presence of 6th order operators is crucial for 
power counting  renormalizability \cite{Horava:2009uw}. On the other hand, in the absence of extra symmetries, radiative corrections will generate all possible terms up to this order~\cite{Sotiriou:2009bx,Blas:2009qj}. In particular, the term $a_ia^i$ in $L_2$ is crucial for the improved dynamical
behavior of the theory~\cite{Blas:2009qj}.

It has been argued that if 
the mass $M_\star$ lies somewhere between roughly 
$10^{10}$GeV and $10^{16}$GeV the theory can both avoid strong coupling, 
and satisfy gravitational
constraints and generic LV constraints in the matter sector \cite{Blas:2010hb} 
(the need to alleviate strong coupling imposes the upper bound \cite{Papazoglou:2009fj,Blas:2009ck} 
which is competing with the lower bounds coming from LV violations, as first pointed out in 
Ref.~\cite{Papazoglou:2009fj}). Thus, at low energy one expects to be able to neglect the 
higher order operators in $L_4$ and $L_6$, so in this sense the action (\ref{SBPSH}) is 
the low energy limit of the action 
(\ref{SBPSHfull}). We can thus say that the low energy or IR 
limit of HL gravity is equivalent to \ae-theory with a hypersurface orthogonal aether.

It is worth mentioning that, since $L_4$ and $L_6$ contain higher spatial
derivatives of the fields, a theory described by action
(\ref{SBPSHfull}) can have solutions that are perturbatively far from
solutions of the theory described by action (\ref{SBPSH}) and which diverge as $M_\star$ goes to infinity, if the
derivatives of the fields are large enough.
That is, theory (\ref{SBPSHfull}) can have more solutions than theory
(\ref{SBPSH}) when contributions coming from $L_4$ and $L_6$
are important.

\section{Static, spherically symmetric, asymptotically flat, regular black holes}

\label{spherical}

\subsection{Horizons and field redefinition}

\label{hred}

In this section we briefly review the field and coupling constant redefinitions that we perform in order to simplify 
our calculations.
Such redefinitions closely follow those used in Ref.~\cite{Eling:2006ec}, to which we refer for a more detailed discussion.

\AE-theory possesses spin-2, spin-1 and spin-0 propagating degrees of freedom, whereas HL gravity has only spin-2 and spin-0 modes. 
However, once spherical symmetry has been imposed, only the 
spin-0 mode is relevant.
The squared speed (at low energies for HL gravity) of this mode, 
defined relative to the aether rest frame, is \cite{Jacobson:2004ts}
\beq
\label{speeds}
s_0^2=\frac{c_{123} (2-c_{14})}{c_{14}(1-c_{13})(2+c_{13}+3 c_2)}\,.
\eeq
Since different modes propagate at different speeds there will be multiple (causal) horizons. In fact for each of these modes the corresponding horizon will be a null surface of the effective metric
\beq
\label{effmet}
g^{(i)}_{\a\b}=g_{\a\b}+(s_i^2-1)u_\a u_b\,, 
\eeq
where $s_i$ is the speed of the spin-$i$ mode.
See Ref.~\cite{Eling:2006ec} for a more detailed discussion.

The
action (\ref{S}) is 
invariant under the combined metric and aether field redefinition
\bea
\label{redmet}
&& g'_{\a\b}=g_{\a\b}+(\sigma-1)u_\a u_\b\,,\\
&& u'^\a=\frac{1}{\sqrt{\sigma}} u^\a\,,
\eea
provided that the $c_i$ are replaced by new parameters $\tilde{c}_i$ which are
functions of the initial $c_i$ (see Ref.~\cite{Foster:2005ec} for the exact correspondence). 
By
choosing $\sigma=s_0^2$ we can make the spin-0 horizon coincide with 
the metric horizon of the redefined metric. 
This will help to simplify the calculations.

As explained in the previous section,
when spherical symmetry is imposed the aether is hypersurface orthogonal,
and so it has vanishing twist. Thus by making use of Eq. (\ref{twist}) 
it is possible to set 
$c_4$ to zero without loss of generality. 
It is worth stressing that this procedure has to come after the field redefinition  
described previously to make the spin-0 horizon coincide with the metric horizon. 
If the order were to be reversed, the field redefinition would regenerate
a $c_4$ term.

\subsection{Asymptotics, regularity and parameters of solutions}
\label{asymreg}

We find it convenient to work in Eddington-Finkelstein-like coordinates with the line element
\beq
\label{efmetric}
ds^2=F(r)dv^2-2 B(r)dv dr-r^2d\Omega^2\,,
\eeq
as these coordinates are regular at both the metric and the spin-0 horizon,
as well as in the interior region of the black hole. We stress that from Eq.~\eqref{efmetric} it follows that the radial coordinate has a geometric meaning,
namely $4\pi r^2$ is the area of a symmetry sphere.
The aether field can be written in the form
\beq
u^\a \partial_\a=A(r) \partial_v-\frac{1-F(r) A^2(r)}{2 B(r) A(r)}\,\partial_r\,,
\eeq
where we have 
imposed
the unit constraint. 
For static, spherically symmetric solutions these  
ans\"atze 
for the metric and the aether can be adopted without loss of generality.
Surfaces of constant $v$ are null, and if we think of 
$v$ as increasing in the future direction
at fixed $r$, it is an ingoing null coordinate. The timelike vector $u$ is 
then future pointing provided $A>0$. The Lagrangian (\ref{Lae}) is even in 
$u$, hence in any solution we can also replace $u$ by $-u$, which 
amounts to replacing $A$ by $-A$. Thus, a black hole with aether 
flowing in can also be viewed as a white hole with aether flowing 
out.

In GR, according to Birkhoff's theorem, 
there is a one parameter family of spherically symmetric solutions, 
namely the Schwarzschild solutions, labeled by the mass. 
These solutions are asymptotically flat, and static. In \ae-theory there
is a scalar mode, corresponding to radial tilting of the aether,
so that Birkhoff's theorem does not apply. Not only are spherical solutions 
not generally static, but even if we restrict to static, spherical solutions, 
they are not necessarily asymptotically flat. In fact, as shown in 
Ref.~\cite{Eling:2006ec}, there is a 
three parameter family of such solutions, and imposing asymptotic 
flatness reduces this to a two parameter family. That is, for each mass, there
is a one parameter family. 
As explained below, we will fix this parameter by the condition that 
the spin-0 horizon, {\em i.e.}~the outermost trapped surface for spin-0 waves,
be nonsingular.

A series expansion for the static, spherically symmetric, asymptotically flat
solutions was previously found
for the case $c_3=-c_1$~\cite{Eling:2006ec},
and we have found that this series remains valid
with no restriction on $c_3$. The result,
given in terms of the inverse radial
coordinate $x=1/r$, is
\begin{eqnarray}
F(x) &=& 1+F_1x+\frac{1}{48} c_{14} F_1^3 x^3+ \cdots \label{asyF}\\
B(x) &=& 1+\frac{1}{16} c_{14} F_1^2 x^2-\frac{1}{12}c_{14} F_1^3
x^3+\cdots \label{asyB} \\
A(x) &=& 1-\half F_1 x+A_2 x^2+\nonumber\\
&&\left(-\frac{1}{96} c_{14}
F_1^3+\frac{1}{16} F_1^3-F_1 A_2\right) x^3+\cdots \label{asyA}
\end{eqnarray}
where $F_1=F'(x=0)$ and $A_2 = A''(x=0)$,
and where $v$ is scaled to set $F(x=0)=1$. 
No more free parameters seem to appear at higher orders, so
the asymptotically flat solutions are determined by
the two free parameters $F_1$ and $A_2$.
The parameters
$c_2$ and $c_{3}$ enter at higher orders in $1/r$. Their absence
at lower orders is presumably related to the fact that they do not appear {\it at all} in solutions
whose aether is everywhere aligned with the timelike Killing vector~\cite{Eling:2006df}~.

We will restrict attention in this paper to  
the values of the coupling coefficients $c_i$ in the
action for which 
presently known 
observational bounds are all met. Even though section \ref{parameterspace} is devoted to a discussion 
about these bounds and the restrictions they will bring to the parameter space, we anticipate 
here
the discussion of a particular constraint: the vacuum \v{C}erenkov constraint \cite{Elliott:2005va}, which amounts 
to the requirement that the speed of the spin-0 mode be greater than or
equal to the speed of light 
($c=1$)
defined by the metric cone.
This means that the spin-0 horizon lies inside the metric horizon.
If matter is minimally coupled to the metric
then the matter cannot propagate beyond the metric cone, and the
spin-0 horizon is hidden from view, except by gravitational and aether signals.
Nevertheless, we shall require that the spin-0 horizon be regular,
simply because the evidence we have  indicates that when a black
hole forms in a collapse process, the spin-0 horizon is in fact
nonsingular. This evidence amounts to the argument that
nothing singular is happening in the fields at that point, bolstered
by numerical simulations for a few examples \cite{Garfinkle:2007bk},
although a general proof has not been given.
If instead collapse does {\it not} automatically impose this condition, 
a black hole would have ``hair" determined by the parameter $A_2$
in Eq.~(\ref{asyA}).

Having imposed asymptotic flatness, and a regular spin-0 horizon, 
there remains a one parameter family of solutions, \textit{i.e.} one solution for each value of the total
mass. Equivalently, the solutions can be parametrized by the horizon radius.
If we adopt units for the radial coordinate in which the horizon radius
is unity, this leaves a unique solution. This is the solution we are characterizing 
in the present paper, as a function of the couplings $c_i$.

\subsection{Equations and constraints}

In this section we discuss the set of equations to be integrated. 
The equations to be solved are the generalized Einstein equations (\ref{gfe}), 
for which we introduce the notation
\beq\label{Emn}
E^{\m\n}\equiv G^{\m\n}-T_{\ae}^{\m\n}=0\,,
\eeq
and the aether field equations (\ref{AEeq}), for which we introduce the notation
\beq
\AE^\m=0\,.
\eeq
Given the spherical symmetry and staticity, many of these
equations are redundant or trivial, and it suffices to 
impose 
\beq\label{eqns}
E^{vv}=E^{vr}=E^{rr}=E^{\theta \theta}=\AE^v=0,
\eeq
each of which must hold at every value of $r$. 
(In particular, $\AE^r$ is proportional to $\AE^v$
so need not be separately imposed.)
All these
equations involve second derivatives of $F$ and $A$ (but not $B$)
with respect to $r$.
However, note that there are only three functions to be solved for, 
$F$, $B$, and $A$, so only three equations are needed to determine
a solution, given initial data. In fact, among these five equations,
two independent combinations are 
initial value constraint equations, relating the functions and their
first derivatives. The constraint equations automatically hold at all values of $r$,
if they are imposed at one value of $r$, as a consequence of the
remaining ``evolution''  equations.

To clarify the distinction between constraints and evolution equations 
in both \ae-theory and HL gravity, it is
useful to recall how that distinction comes about in GR in the 
general case (\textit{i.e.,} in the absence of symmetries). 
In the absence of matter, the field equations of GR are given by 
the vanishing of the Einstein tensor, 
$G^{\mu\nu}=0$. Because of diffeomorphism invariance,
there are four free functions in the time evolution of the metric, 
which is therefore not uniquely determined
from initial data. This means that not all of the Einstein 
equations can be evolution equations. Indeed, 
the Einstein tensor satisfies the contracted Bianchi identity
\beq
\nabla_\mu G^{\mu\nu}=0\,,
\eeq
which holds independently of any field equations and 
can also be seen as a consequence of the  
diffeomorphism invariance of the 
Einstein--Hilbert action.
Expanding this equation in components in a coordinate system $(t,x^i)$  
one has 
\beq
\label{cbianchiv}
\partial_t G^{t\n} + \partial_i G^{i\n} + \Gamma G\, \mbox{-terms}=0\,.
\eeq
This identity implies that if $G^{\m\n}=0$ at some initial ``time'' $t_0$,
then $G^{t\n}=0$ also holds at $t=t_0+\delta t$.
Thus if the equations $G^{t\n}=0$ are  
imposed at some initial time $t_0$, they are satisfied in the whole spacetime
when the remaining equations hold.
Moreover, the quantities $G^{t\n}$ involve only initial values. This follows from the
fact that if a given field component appears in $G^{\m\n}$ with up to $n$ time derivatives,
then Eq.~(\ref{cbianchiv}) -- being an identity that holds for all metrics independent of field equations --
implies that $G^{t\n}$ has no more than
$n-1$ time derivatives of that field component. Thus the equations
$G^{t\n}=0$ are {\it initial value constraint equations}.

The discussion of initial value constraint equations can be 
extended to cover the case when matter is coupled to 
the metric. We will discuss that extension explicitly here
just for the case when the ``matter'' corresponds to the 
aether degrees of freedom in \ae-theory. 
The corresponding Einstein equation
(\ref{Emn}) involves second time
derivatives of the aether field $u^\mu$
in $T_{\ae}^{\m\n}$ [see Eq~\eqref{Tae}]. These arise 
from the variation of the metric in the
Christoffel symbols occurring in the covariant derivatives
of the aether field. The equations
$E^{t\n}=0$ are therefore clearly not initial value
equations, even though the on-shell identity 
$\nabla_\m E^{\m\n}=0$ implies that if they 
hold initially they continue to hold as a result
of the remaining equations.\footnote{Note that
$G^{t\n}-8\pi T^{t\n}=0$
are indeed initial value constraint equations
in settings where the matter stress tensor 
$T^{\m\n}$
has fewer 
derivatives than the matter equations of motion.}
Instead, to identify 
true initial value constraint equations we need
to find a true identity, analogous to the Bianchi identity,
that holds independent of any field equations. 

Such an identity can be found by using the 
diffeomorphism invariance of the full \ae-theory 
action.\footnote{The 
precise origin of this identity, and its generalization to other theories with
tensor matter, will be explained in a forthcoming publication.}
One finds
\beq\label{identity}
\nabla_\m (E^{\m\n}-u^\mu\AE^\n)= \AE_\m\nabla^\n u^\m,
\eeq
where the normalization of the aether field
equation is defined by $\d S/\d u^\m = 2\AE_\m$.
This identity can be used to argue, in a way 
identical to that used for vacuum GR, that 
\beq\label{constraint0}
E^{t\n}-u^t\AE^\n=0
\eeq
are initial value 
constraint equations. 

The reasoning presented above
applies as well to the static, spherically symmetric case,
with the role of $t$-evolution replaced by $r$-evolution.
Hence we now define the constraint equations as
\beq\label{constraint}
C^\nu\equiv E^{r\n}-u^r\AE^\n=0\,.
\eeq
It follows from the reasoning just given that, 
once imposed at a single initial radius, 
these equations are automatically satisfied 
at all $r$ provided that the
remaining field equations hold. Also, they 
depend only on initial data (with respect to $r$-evolution).
To exploit this structure, we therefore replace the set of equations
(\ref{eqns}) by the equivalent set
\beq\label{eqns2}
E^{vv}=E^{\theta \theta}=\AE^v=0,\quad C^v=C^r=0\,.
\eeq
The first three equations can be recast in the 
form
\bea
\label{Fpp}
F''&=& F''( A, A', B, F, F')\label{ev1}\\
\label{App}
A''&=& A''( A, A', F, F')\label{ev2}\\
\label{Bp}
B'&=&B'( A, A', B, F, F')\label{ev3},
\eea
which is a system of ordinary differential equations (ODEs) that
can be numerically integrated with respect to $r$.
These equations will be 
our ``evolution'' equations. The constraint equations $C^v=C^r=0$, 
instead, depend only on $A$, $A'$, $B$, $F$ and $F'$.
Therefore they simply impose algebraic constraints on the 
data at the ``initial'' radius $r_0$, and are automatically
preserved by the evolution equations for any other $r$.

\subsection{Numerical implementation}
\label{code}

As explained in section \ref{asymreg}, imposing asymptotic flatness, 
together with the condition that  
there is a regular spin-0 horizon,
will lead to a 
one parameter family of solutions. 
These describe black holes, because for the theory parameters that we consider 
there will be a metric horizon outside the spin-0 
horizon.
We begin the integration at the location of the spin-0 horizon, where
the regularity condition can be imposed directly on the initial
data, as was done in Ref.~\cite{Eling:2006ec}.
To conveniently implement this procedure, as in Ref.~\cite{Eling:2006ec}, 
we first make a field redefinition so that the spin-0 
metric, Eq.~(\ref{effmet}) with $s_i=s_0$, 
is the new metric. (This induces a change in the
coupling parameters, which we keep track of.) Then the spin-0 horizon 
coincides with the metric horizon at $r=r_H$, 
which is defined by the condition
\beq
\label{constraint1} F(r_H)=0\,.
\eeq
We adopt units in which $r_H=1$, so the one-parameter family is
represented with just a single solution to be found. 
After finding the solution, we then make the inverse 
of the 
field redefinition (\ref{redmet})
to express the solution in terms of the original metric that
is assumed to be minimally coupled to the matter fields,
and the corresponding aether field. 

To impose regularity of the horizon we proceed as follows. 
The evolution equation for $B'$ [Eq. (\ref{Bp})] turns out to have the structure
\beq\label{BpBis}
B'= b_0/F + b_1 + b_2F, 
\eeq
where 
$b_{0,1,2}$ are functions of $(A,A',B,F')$. Therefore, 
$B'$ diverges at $r_H$ unless 
\beq \label{constraint2}
b_0(A,A',B,F')|_H=0\,.
\eeq 
(Here and below, we shall 
denote quantities evaluated at $r_H$ using the subscript $H$, {\it e.g.}\ 
$F_H\equiv F(r_H)$.)
Once the constraint in Eq.~\eqref{constraint2} 
is imposed, both the metric and aether are regular at the horizon. 

The system of evolution
equations 
(\ref{Fpp})--(\ref{Bp}) 
requires a five dimensional space of
initial conditions, $(A,A',B,F,F')_H$, 
but we also have to impose 
the constraint equations $C^r=C^v=0$. 
Once we impose the horizon and regularity conditions
Eqs.~\eqref{constraint1} and \eqref{constraint2}, 
the constraint equation $C^r=0$ is automatically
satisfied at $r_H$, as it is proportional to $F$. 
The other constraint equation $C^v=0$, however, 
is not trivially satisfied at $r_H$, and 
further restricts the initial conditions:
\beq \label{constraint3}
C^v(A,A',B,F')|_H=0\,.
\eeq 
Together with the other conditions
in Eqs.~\eqref{constraint1} and \eqref{constraint2}, 
this cuts the space of initial conditions from five down to two
dimensions. This is further reduced to one dimension 
by choosing to scale the 
coordinate $v$ so as to have 
$B_H=1$. We can then parametrize the space of initial data with the value of 
$A$ on the horizon, $A_H$. 

A generic value of $A_H$ will not lead to an asymptotically flat solution.
As in Ref.~\cite{Eling:2006ec}, 
we seek the value of  $A_H$ leading to asymptotic flatness 
using a ``shooting method''.
In practice, we integrate out from $r_H$ 
starting with different values of  $A_H$, until we find both a value 
of $A_H$ that gives $F A^2>1$ far away from the horizon, and 
one which instead gives $F A^2<1$. This is sufficient to ``bracket''
the asymptotically flat solution, which as can be seen from Eqs.~\eqref{asyF}--\eqref{asyA} satisfies $\lim_{r\to\infty} FA^2=1$. (Note that the quantity
$FA^2$ is invariant under rescalings of $r$ and $v$, which justifies its use in our code, where we choose specific scalings for these coordinates).
Once the two bracketing values of $A_H$ have been identified, a simple bisection procedure will yield the asymptotically flat black hole with higher
and higher accuracy.

In principle this completes the description of our integration procedure. However
there are some complications that affect how it is actually implemented in our code. 
We now describe these complications and the implementation.

First, the value of $A_H$ does not uniquely fix all the initial
conditions through Eqs.~\eqref{constraint1}, \eqref{constraint2} and \eqref{constraint3}, 
since the latter two are quadratic in $F'_H$ and $A'_H$. 
These can be linearly combined to obtain a linear equation for 
$F'_H$, but when replacing the solution
into Eq.~\eqref{constraint2} or~\eqref{constraint3} one obtains a quartic equation in $A'_H$. Thus one $A_H$ determines four values for $A'_H$ (and $F'_H$, which depends on $A'_H$).

Since $F$ must asymptote to a positive value at spatial infinity to achieve asymptotic flatness, we can readily discard
the branches that give $F'_H<0$, otherwise $F$ would then have another 
zero (\textit{i.e.,} another horizon, possibly singular) outside $r_H$.
However, this  still does not select a unique branch, and we are typically 
left with at least two branches that can potentially
give an asymptotically flat black hole with no horizons outside  $r_H$.
We find, however, that 
for any given set of 
the theory's parameters, only one branch seems 
to give rise to an  asymptotically flat black hole,
the other branches failing to give a viable
bracketing interval for $A_H$. 
Because the branch that gives the asymptotically flat solution 
is typically not the same (as labeled by Mathematica)
as the parameters of the theory are varied, in practice we proceed in the following manner. We start off with a theory parametrically close to GR, where
we can easily identify the branch that evolves to an asymptotically flat solution,
that being the branch that gives a value of $F'_H$ close to the GR value
$F'_H=1/r_H=1$. [The other branches give instead very large values for $F'_H$, 
and they fail at providing a bracketing interval for $A(r_H)$.]
We then gradually move away from GR, varying the coupling parameters 
$c_i$ by a small amount, 
and identifying the branch that is closest to the one that worked for 
the previous values of $c_i$. While this procedure might in principle 
miss the existence
of another branch of asymptotically flat black hole solutions, 
that seems unlikely in view of the tests that we conducted on the 
remaining branches, none of which seems to produce such solutions.

Another difficulty stems from the regularity condition \eqref{constraint2}. While
this equation ensures that $B'$ is non-singular on the horizon, Eq.~\eqref{BpBis} can potentially be affected
by numerical inaccuracies when evaluated very close to the horizon. 
This is because both $b_0$ and $F$ are zero on the horizon, 
but are non-zero (albeit small) at radii $r$ very close to $r_H$.
As a result, $B'$ will not be calculated accurately at such radii,
due to the finite machine accuracy of any calculator.
While it might be possible to overcome this problem by cranking up the number of significant digits used by the code, a more elegant and 
robust approach is to integrate the evolution equations perturbatively near the horizon, \textit{i.e.} to expand them in a series in $r-r_H$
and solve them analytically  order by order, as in Ref.~\cite{Eling:2006ec}.
We did so up to seventh order in  $r-r_H$ using Mathematica, 
and used this perturbative solution 
from $r_H=1$ to $r_{\rm in}=1.001$. The error of the 
perturbative solution at $r_{\rm in}=1.001$ is therefore 
${\cal O}(r_{\rm in}-r_H)^8\sim 10^{-24}$,
comparable to the machine accuracy 
(we use 22 significant digits for our real numbers, which is possible using Mathematica).
We can then integrate the evolution equations numerically from 
$r_{\rm in}$ up to very large radii ($r\sim 10^4$ or larger) using Mathematica's default 
ODE integrator (which automatically switches between backward differentiation formulas and Adams multistep methods depending on the equations' stiffness).
Because the evolution equations~\eqref{ev1}--\eqref{ev3} are lengthy and complicated,  we 
evaluate their right-hand sides with 22 significant digits in 
order to minimize the impact of round-off errors, 
and we set the accuracy and the precision
goal of the ODE integrator to $10^{-15}$. As a confirmation that the integration is performed accurately, we check that the 
constraints $C^v$ and $C^r$ are preserved to within $10^{-12}$ or better. More specifically, the 
dimensionless quantities $|C^v r^2|$ and $|C^r r^2|$ remain smaller
than  $10^{-12}$ during the whole evolution. 

Using this set-up, we can then finely bracket the asymptotically flat solution. (We stop our bisection 
either when the value $A(r_H)$ giving an asymptotically flat solution is determined to within $10^{-22}$ or when the two bracketing solutions
agree to within $10^{-15}$.) As final confirmation of the accuracy of our procedure, we rescale the time $v$ 
so that $F$ approaches $1$ asymptotically, and then verify that our solution agrees well with the asymptotic solution \eqref{asyF}--\eqref{asyA}.

\section{Parameter space}
\label{parameterspace}

Even after the parameter redefinition to eliminate $c_4$ in \ae-theory described in section \ref{background}, one is left with a 3-dimensional parameter space. Similarly, in HL gravity one has to deal with a 3-dimensional parameter space {\em a priori}. Scanning such a space is clearly a formidable task. Fortunately, there are certain regions of the parameter space which are far more interesting than others, as there are a number of viability constraints that one can impose on both theories, namely: 
\begin{enumerate} 
\item {\it Classical and quantum-mechanical stability}: 
all propagating modes should be classically stable and have positive energy (no tachyons, no ghosts).
\item {\it Avoidance of vacuum  \v Cerenkov radiation by matter} \cite{Elliott:2005va}: this requires that the squared speeds of the propagating modes should be greater than 
or equal to unity.
\item {\it Agreement with 
GR at first post-Newtonian order}:
(This implies, in particular, that all the constraints coming from
Solar system experiments are met.)
The parametrized post-Newtonian (PPN) parameters of both 
\ae-theory and low-energy HL gravity are identical to those of 
GR with the exception of those measuring preferred frame effects, $\alpha_1$ and $\alpha_2$ \cite{Eling:2003rd,Blas:2010hb}.  
These 
two parameters are constrained to be below $10^{-4}$ and $10^{-7}$ respectively \cite{Will:2005va}.
\end{enumerate}
Note that constraints 2 and 3 refer implicitly to the metric to which matter couples
minimally. Hence we must impose these constraints on the coupling parameters 
{\it before} making the field redefinition (\ref{redmet}). 
In what comes next we explore only the part of the parameter space that satisfies all of the above constraints. Additionally, we impose a stronger version of 3, 
namely
$\a_1=\a_2=0$, 
so
that the two theories are indistinguishable from GR at the 
first order 
PPN level. (This requirement is reasonable given that
the bounds on $\alpha_1$ and $\alpha_2$ are very strong, as mentioned above.)

The bound imposed on the $c_i$ by the above constraints have been summarized in Ref.~\cite{Jacobson:2008aj} for \ae-theory. The condition $\a_1=\a_2=0$ translates to
\bea
&&c_2=\frac{-2 c_1^2-c_1 c_3+c_3^2}{3 c_1}\,,\\
&& c_4=-\frac{c_3^2}{c_1}\,,
\eea
which reduces
the parameter space down to 2 dimensions. 
In terms of 
$c_{\pm} =c_1\pm c_3$,
then constraints 1 and 2 are satisfied in the region
\bea
&&0\leq c_+\leq 1\,,\\
&&0\leq c_-\leq \frac{c_+}{3(1-c_+)}\,.
\eea
For practical purposes, and given that larger values are unlikely to be compatible with strong field
constraints from binary pulsar systems~\cite{Foster:2007gr}, we will explore the part of this region which also satisfies $c_-\leq1$.

An important observational constraint that has not been included in the 
list above is that related to gravitational radiation 
from binary pulsars. As shown in Refs.~\cite{Foster:2006az,Foster:2007gr}, 
when the preferred frame PPN parameters $\a_1$ and $\a_2$ vanish, 
as assumed here, and when gravitational fields are 
everywhere
weak or
the coupling constants $c_i$ are smaller than something of order $\sim 0.01-0.1$,
all gravitational radiation is sourced by the quadrupole $Q_{ij}$, as in GR.  
The net power radiated in all modes is then
given by 
$(G_N {\cal A}(c_i)/5) \dddot{Q}^2_{ij}$. 
Agreement with the damping rate of 
GR requires ${\cal A}(c_i)=1$, which would impose an extra
relation between the $c_i$. Even though we have not used this constraint to further 
restrict the parameter space, we 
present the curve ${\cal A}(c_i)=1$ in some of the figures that follow,
but  only for $c_+<0.1$, as for larger values the damping rate for compact
binaries is not accurately given by this formula~\cite{Foster:2007gr}.

We now move to HL gravity. The PPN constraints were worked out in Ref.~\cite{Blas:2010hb}, but a different parametrization $(\mu,\a,\b)$ of the action was used (what we denote here as $\mu$ was actually denoted by $\lambda'$ in Ref.~\cite{Blas:2010hb}). The relation with the more common parametrization used in Eq.~(\ref{SBPSH}) is given by
\beq
\label{HLtoblas}
\xi=\frac{1}{1-\beta}, \quad \lambda=\frac{1+\mu}{1-\beta},\quad \eta=\frac{\alpha}{1-\beta}\,.
\eeq
The relation with the \ae-theory parameters, given by  (\ref{HLpar}), is
\beq
\label{blastoae}
\a = c_{14}, \quad \b = c_{13}, \quad \mu=c_2\, .
\eeq
The condition $\a_1=\a_2=0$ 
is satisfied when $\a=2 \beta$, 
and again the parameter space becomes 2-dimensional. 
Then constraints 1 and 2 above are satisfied 
in the following 3 regions of the $(\mu,\beta)$ plane
\bea
\label{con1}
&&0<\beta<1/3, \quad \mu > \frac{\beta (\beta+1)}{1-3\beta}\, ,\\\label{con2}
&& 0<\beta<1/3, \quad \mu<- \frac{2+\beta}{3}\, ,\\\label{con3}
&&1/3<\beta<1,\quad \frac{\beta (\beta+1)} {1-3\beta}<\mu<- \frac{2+\beta}{3}\,.
\eea
(Note that the two last regions are actually connected, forming one single region.)
We will explore all 
of these regions but for practical purposes we will restrict to $|\mu|<10$. 

 In terms of the $(\xi,\lambda,\eta)$ set of parameters, the PPN condition $\a=2\b$ becomes $\eta=2(\xi-1)$, and 
Eqs.~(\ref{con1})-(\ref{con3}) 
translate to 
\bea
 \frac{3\lambda-1}{2\lambda}>&\xi&>1,\quad \lambda>1\quad {\rm or} \quad \lambda<0\,,\\
& \xi&>1,\quad 0<\lambda<1/3\,.
\eea

\section{Numerical Solutions}

\label{results}

\subsection{Comparison to known black-hole solutions in \ae-theory}

\begin{table}[b]
\caption{Properties of 
regular black hole solutions with $c_3=c_4=0$ and $c_2$ 
such that $s_0=1$. GR corresponds to $c_1=0$. 
See text for explanation of the quantities shown.} 
% title of Table
\centering % used for centering table
\begin{tabular}{c c c c} % centered columns (4 columns)
\hline\hline %inserts double horizontal lines
$c_1$ & $r_g/r_H$ & $F'_H A_H^2$ & $\gamma_{\rm ff}$ \\ [0.5ex] % inserts table
%heading
\hline % inserts single horizontal line
$0$ & 1 & n/a & n/a\\
$0.1$& 0.989489& 2.09612& 1.60280\\
$0.2$& 0.978021& 2.07168& 1.57695\\
 $0.3$& 0.965229& 2.03920& 1.54768\\
 $0.4$& 0.950547& 1.99652& 1.51409\\
$0.5$& 0.933044& 1.94056& 1.47484\\
 $0.6$& 0.911068& 1.86668& 1.42796\\ 
$0.7$& 0.881313& 1.76732& 1.37024\\
$0.8$& 0.835830&  1.62834& 1.29591\\
$0.9$& 0.747519& 1.41557& 1.19212\\
$0.91$& 0.733012& 1.38702& 1.17904\\
 $0.92$& 0.716505& 1.35637& 1.16523\\
 $0.93$& 0.697454& 1.32324&  1.15060\\
 $0.94$& 0.675075& 1.28711& 1.13502\\
 $0.95$& 0.648165& 1.24724& 1.11831\\ 
$0.96$& 0.614764& 1.20248& 1.10023\\
 $0.97$& 0.571331& 1.15094& 1.08044\\
 $0.98$& 0.510382& 1.08891& 1.05834\\ 
$0.99$& 0.410630& 1.00689& 1.03281\\[1ex] % [1ex] adds vertical space
\hline \hline %inserts single line
\end{tabular}
\label{table:comparison} % is used to refer this table in the text
\end{table}

As a first test of our code, we tried to reproduce the \ae-theory regular black-hole solutions
studied in Ref.~\cite{Eling:2006ec}. In that paper, Eling \& Jacobson focused on \ae-theories
with $c_3=c_4=0$ in order to simplify
the (very complicated) field equations. Additionally, they imposed the condition $c_2=-c_1^3/(3 c_1^2-4 c_1+2)$, which ensures that $s_0=1$, so that the metric
horizon coincides with the spin-0 horizon. After these simplifying assumptions the preferred frame PPN parameter 
$\alpha_1$ vanishes. However, the observational bound on the second preferred frame parameter, $\alpha_2< O(10^{-7})$, leads to the constraint $|c_{1}|,|c_{2}|<O(10^{-7})$. The values for $c_1$ considered in Ref.~\cite{Eling:2006ec} where actually significantly larger, making the theories considered there non-viable.
(Even if one identifies the theories studied in Ref.~\cite{Eling:2006ec} with the ``redefined'' theory of Sec.~\ref{hred}, they do not lead to 
viable theories in the ``physical'' parameter space. 
One should mention, however, that the purpose of Ref.~\cite{Eling:2006ec}
was to understand the properties of regular \ae-theory black holes in a simple family of ``test-theories'', 
even at the cost of sacrificing their viability.)

Ref.~\cite{Eling:2006ec} focused on theories with $c_1>0$ (since $c_1<0$ would give negative spin-0 mode energy), 
but found no regular solutions for $c_1>0.7$. 
Our code, instead, finds regular black-hole solutions
in the whole range $0< c_1<1$. While it is not clear why the solutions for $0.7<c_1<1$ 
were not found by the code of Ref.~\cite{Eling:2006ec} --- 
nor 
why they were not produced in the gravitational-collapse simulations of Ref.~\cite{Garfinkle:2007bk}, 
which confirmed that no regular BHs
seemed to form in theories with $c_1>0.8$ --- our code produces, even in this region of the parameter space, regular BHs 
that are very accurate solutions of the field equations and that
agree with the asymptotically flat analytical solution~\eqref{asyF}--\eqref{asyA} (see Sec. \ref{code}).

\begin{figure}
\includegraphics[type=pdf,ext=.pdf,read=.pdf,width=8.5cm]{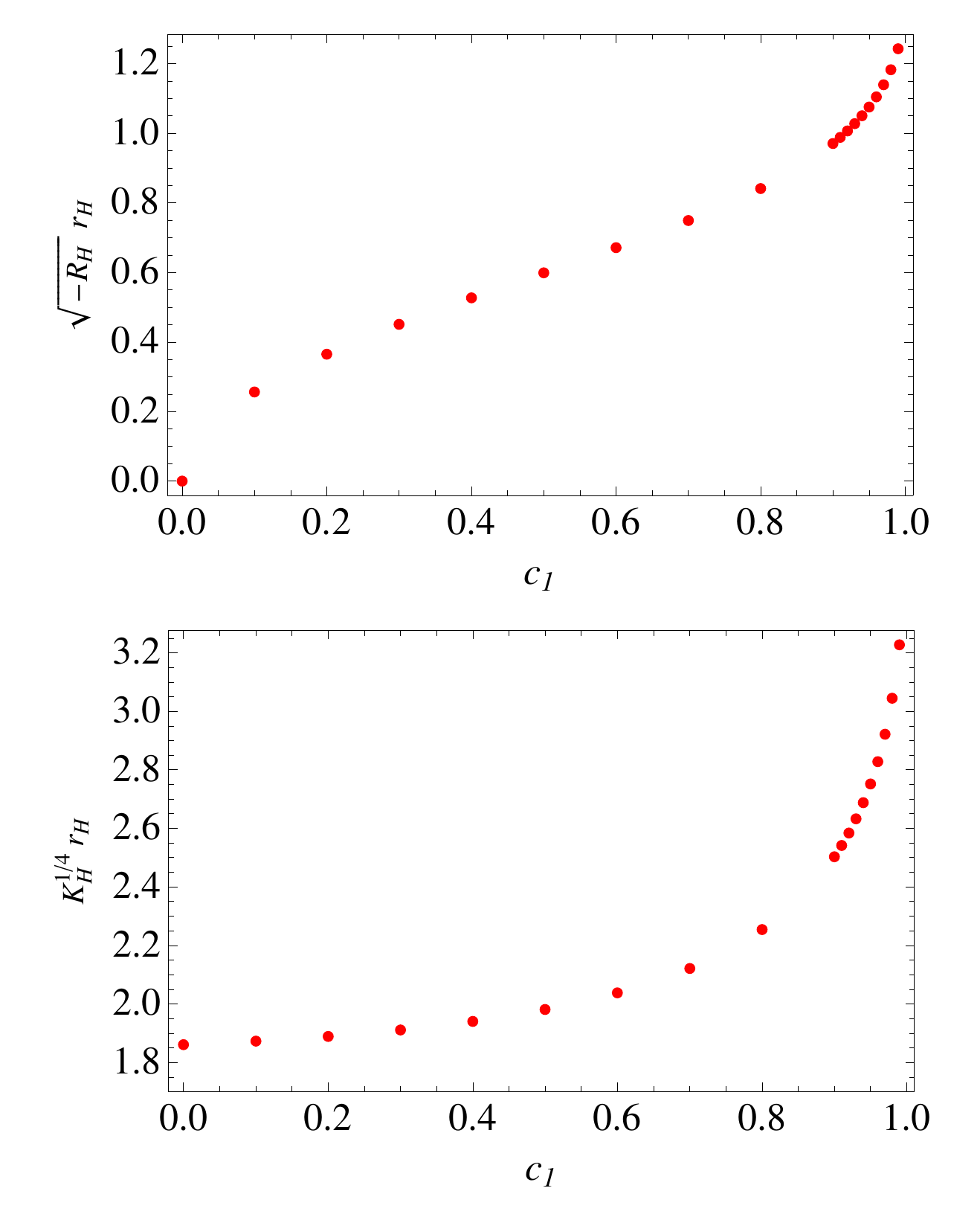}
\caption{\label{curvature_figure} The curvature radius on the horizon, in units of the horizon's circumference radius $r_H$.
More specifically, $R_H$ and $K_H$  denote the Ricci and  Kretschmann scalars evaluated on the horizon.}
\end{figure}

One possibility is that the code of Ref.~\cite{Eling:2006ec} was simply not accurate enough: because of the lengthy field equations,
rounding errors can propagate and lead to significant inaccuracies if one does not use a sufficient number of significant digits.
(For this work, as stressed in Sec.  \ref{code}, we use real numbers with 22 significant digits). 
 To gauge the accuracy of the code of Ref.~\cite{Eling:2006ec}, we 
compare the numerical solutions found there with those that we find with our code. 
In particular we look at three quantities that were used in Ref.~\cite{Eling:2006ec} to characterize the solutions,
namely (i) the 
ratio
$r_g/r_H$ (which equals 1 in GR),
where $r_H$ is defined geometrically as the 
proper 
circumference of the horizon
divided by $2 \pi$,
and where
 $r_g$ is the ``gravitational radius,'' {\it i.e.} the parameter that
appears in the asymptotic form of the metric, 
$F = 1 - r_g/r +O(1/r^2)$;
in terms of the mass $M_{\rm tot}$ as measured by a distant observer we have
\beq\label{gravrad}
r_g=2 G_N M_{\rm tot};
\eeq
(ii) the combination
$F'_H A_H^2$, which
is invariant under a rescaling of the coordinate time
$v\to \chi v$; (iii)  the Lorentz factor
$\gamma_{\rm ff}=u^\mu u_\mu^{\rm obs}$
of the aether, defined with respect to a unit Killing energy radial
observer at the horizon ($u_\mu^{\rm obs}$ is tangent to 
a radial free-fall trajectory that starts at rest at spatial infinity).

\begin{figure}
\includegraphics[type=pdf,ext=.pdf,read=.pdf,width=8.5cm]{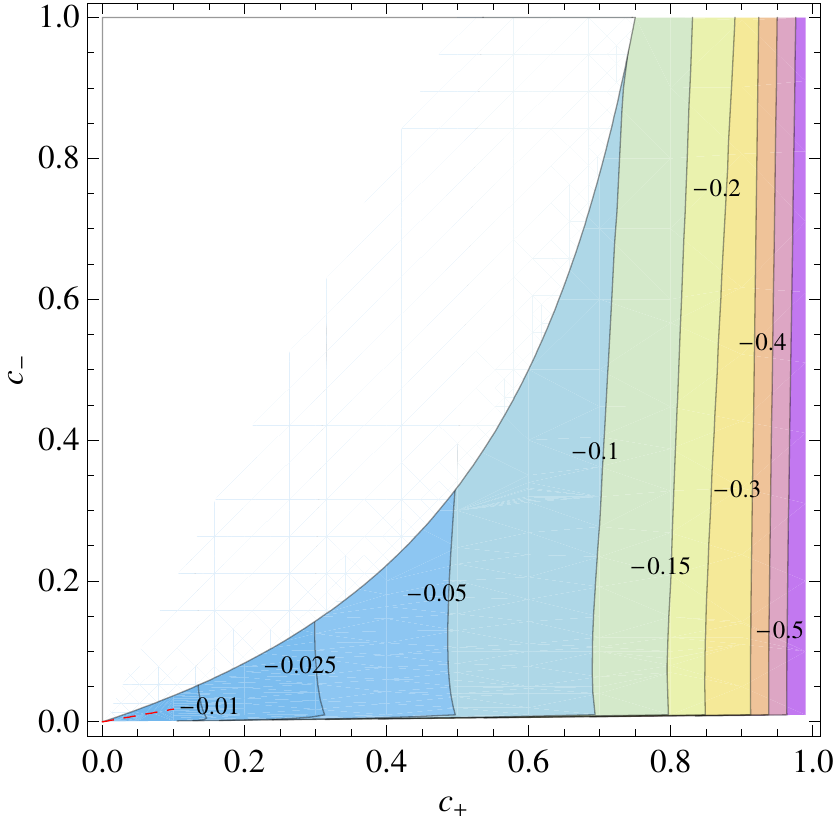}
\caption{\label{wiscoAE} Fractional 
\ae-theory deviation from GR  for  the dimensionless product $\omega_{_{\rm ISCO}}r_g$ of the
ISCO frequency and the gravitational radius,
in the viable region of the parameter plane (see Sec.~\ref{parameterspace}).
The red dashed line extending up to $c_+\approx0.1$ 
is the binary pulsar constraint $c_-\approx 0.18 c_+$.
}
\end{figure}

 \begin{figure}
 \includegraphics[type=pdf,ext=.pdf,read=.pdf,width=8.5cm]{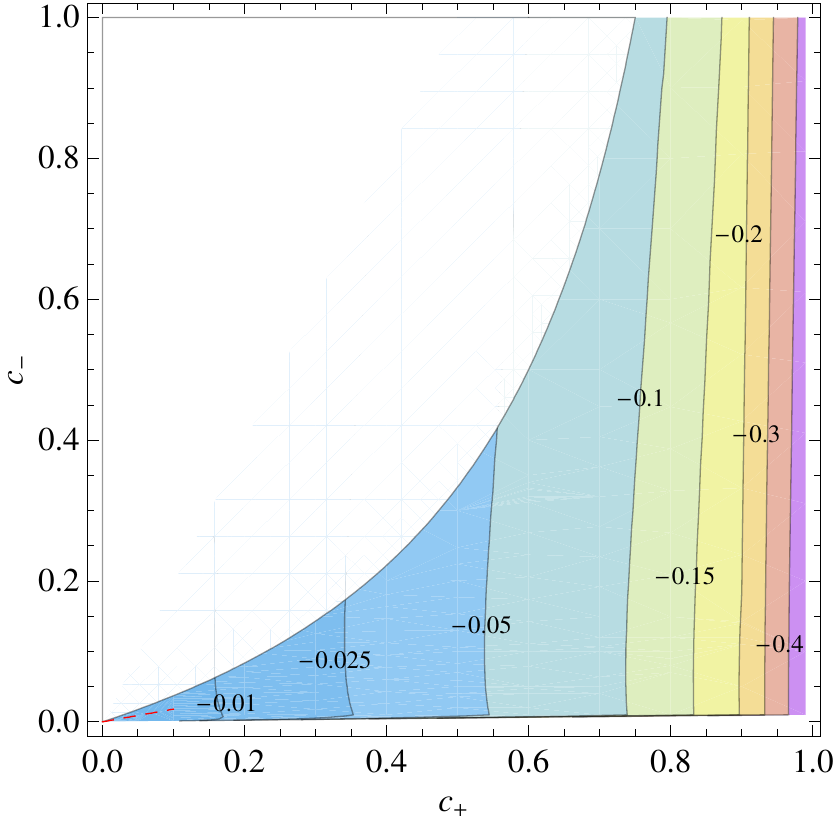}
 \caption{\label{zmaxAE} Same as Fig.~\ref{wiscoAE}, but for 
 the maximum redshift of a photon emitted by a source moving on the ISCO, 
$z_{\rm max}=\nu_{\rm emitted}/\nu_{\rm measured}-1$.
This maximum redshift is 
obtained for a photon emitted ``backward'' with respect to the velocity of the source.
 }
 \end{figure}

These quantities are reported in Table I as function of $c_1$ (GR corresponds to $c_1=0$), for the solutions that we
find in the range $0<c_1<1$. A comparison with Table I of Ref.~\cite{Eling:2006ec} shows that the values of $r_g/r_H$
only agree up to the third decimal digit for $c_1\leq0.3$. This difference 
is much larger than our estimated error on  $r_g/r_H$, which is less than 
$2$ parts in $10^{14}$ and which we obtain from the bracketing 
procedure described in Sec. \ref{code}. 
This suggests that our code might be more accurate than that of  Ref.~\cite{Eling:2006ec}, 
which could explain why we find solutions up to $c_1=0.99$.\footnote{Our values
for $\gamma_{\rm ff}$ differ significantly from those reported in 
Ref.~\cite{Eling:2006ec}, but we have determined that the latter were
incorrectly computed without accounting for the non-standard normalization
of the metric function $F(r)$ at infinity.}

Although we find regular black holes up to $c_1=0.99$,
we too do not find solutions for $c_1\geq1$. 
There is some evidence that the would-be horizon becomes singular
in this case. 
Indeed, when $c_1$ approaches 1, the Ricci scalar $R$ and the Kretschmann scalar
$K=R^{\alpha\beta\mu\nu}R_{\alpha\beta\mu\nu}$ evaluated on the 
horizon grow rapidly, as shown in Fig.~\ref{curvature_figure},
and at least $K_H$ appears to be diverging, 
suggesting
that no regular black-hole solutions exist when $c_1=1$.

Lastly let us note that, as expected, the dimensionless ratio $r_g/r_H$ goes to the GR value ($1$) when
$c_1$ is small, and decreases for larger $c_1$.
The deviations away from GR can be 
very significant as can be seen from Table I, but this is
only because the theories under consideration here do not satisfy the constraints of Sec. \ref{parameterspace}
and are therefore allowed to deviate
significantly from GR. As we will see in the next section, theories which satisfy all the constraints available to date 
only allow regular BH solutions that
are very similar to the Schwarzschild BHs of GR.

\begin{figure}
\includegraphics[type=pdf,ext=.pdf,read=.pdf,width=8.5cm]{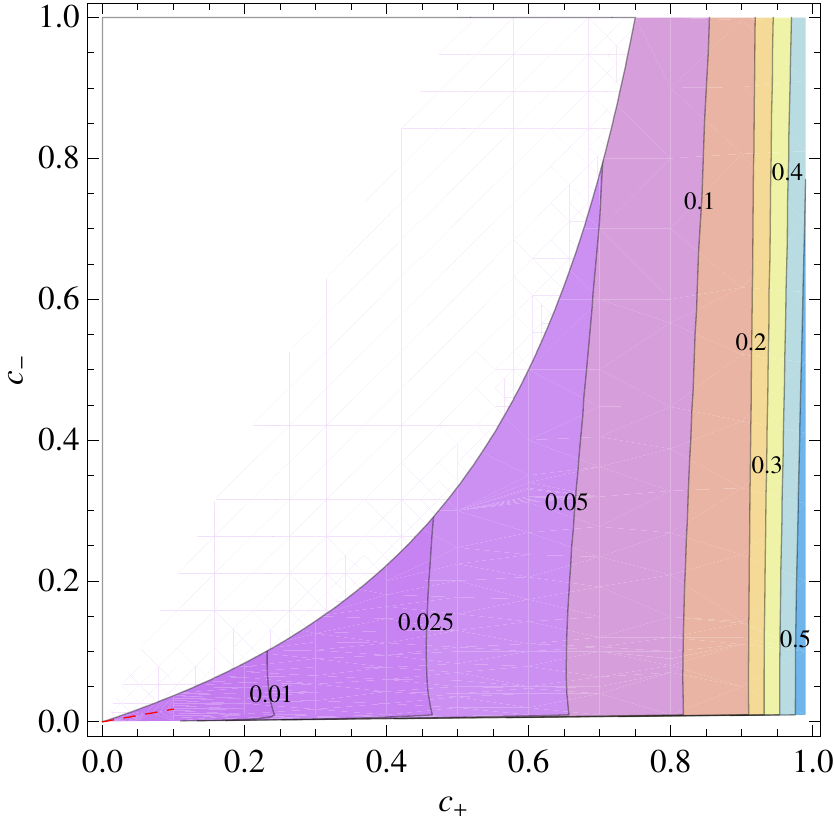}
\caption{\label{bphAE} Same as Fig.~\ref{wiscoAE}, but for the 
impact parameter of the circular photon orbit $b_{\rm ph}$ in \ae-theory,
normalized against  the gravitational radius $r_g$.
Note that the frequency of the circular photon orbit is $\omega_{\rm ph}=1/b_{\rm ph}$.
}
\end{figure}

\begin{figure}
\includegraphics[type=pdf,ext=.pdf,read=.pdf,width=8.5cm]{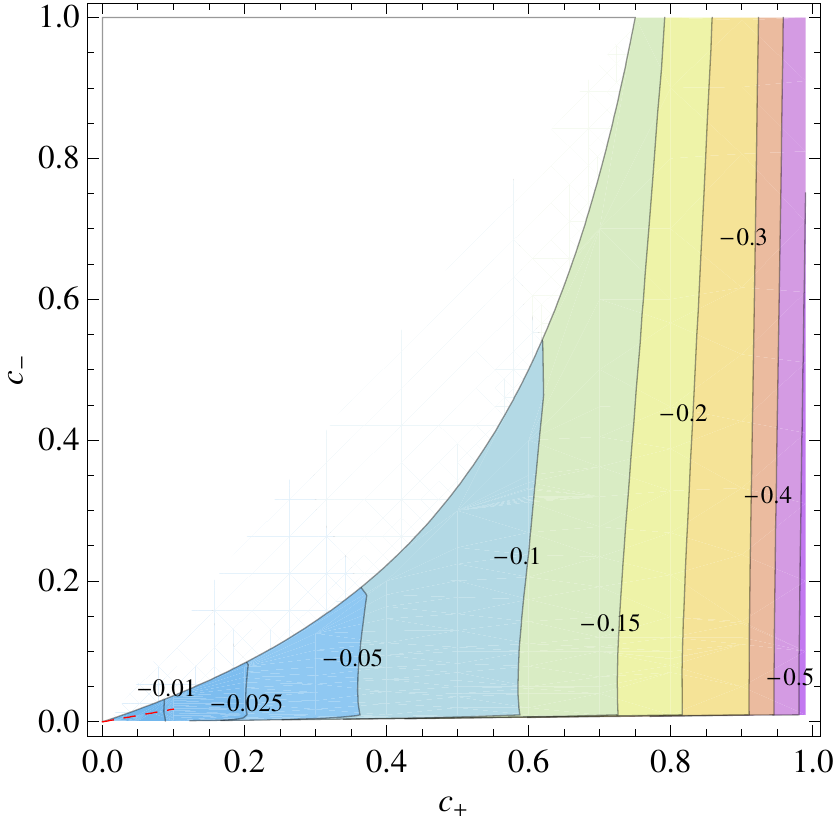}
\caption{\label{massAE} Same as Fig.~\ref{wiscoAE}, but for the 
ratio
 $r_g/r_{H}$  of the gravitational radius $r_g$ to the horizon's circumference radius $r_H$.}
\end{figure}

\subsection{Viable regular BHs in \ae-theory and HL gravity}

In this section we focus on the regions of 
the parameter spaces of \ae-theory and HL gravity that
satisfy all the 
observational constraints available to date, as described in Sec.~\ref{parameterspace}.
To this purpose, we have considered 236  points in the parameter plane  $(c_{+},c_{-})$ of \ae-theory and
405 in the parameter plane  $(\beta,\mu)$ of HL gravity. Each of these points corresponds to a different gravity theory,
and for each of them we have derived the regular asymptotically flat spherical BH solution as described in Sec.~\ref{code}.
To characterize the solutions, we have then extracted, for each of them, the following quantities (explicit expressions for which can be found in the Appendix):
\begin{enumerate}
\item The dimensionless 
product $\omega_{_{\rm ISCO}}r_g$: 
$\omega_{_{\rm ISCO}}$ is the orbital frequency of the 
innermost stable circular orbit 
(ISCO), while $r_g$ is 
the gravitational radius (\ref{gravrad}). 
The deviation of this quantity away from its GR value ($2\cdot6^{-3/2}$) is a measure of how easily \ae-theory (or 
HL gravity) might be discriminated
from GR using the X-ray continuum spectra of accretion disks~\cite{zhang,mcclintock} (see also Ref.~\cite{me_cosimo} on how to use these spectra to detect generic 
deviations from GR black holes) or using future gravitational-wave observations of stellar-mass black holes orbiting around a supermassive black hole (these sources
are known as extreme mass-ratio inspirals, or EMRIs)~\cite{emris_review}.
\item The maximum redshift (measured at spatial infinity) for a photon emitted by a source moving on the ISCO, 
$z_{\rm max}=\nu_{\rm emitted}/\nu_{\rm measured}-1$. The maximum redshift comprises both the
redshift due to gravitational field of the black hole, and the Doppler redshift, which is
maximum for a photon emitted ``backward'' (\textit{i.e.,} in the negative $\phi$ direction, if the source moves 
in the positive $\phi$ direction). The deviation of this quantity away from its GR value ($3/\sqrt{2} - 1$), together with the value
of the combination $\omega_{_{\rm ISCO}}r_g$ mentioned above, 
is a measure of how easily a black hole in \ae-theory (or 
HL gravity) could be distinguished
from a Schwarzschild black hole using iron-K$\alpha$ lines~\cite{fabian,chris} (see also Ref.~\cite{psaltis} on how to use iron-K$\alpha$ lines
to detect deviations from GR black holes).
\item  The dimensionless 
ratio $b_{\rm ph}/r_g$, where $b_{\rm ph}$ is
the impact parameter of the circular photon orbit. The deviations of this quantity
from GR (where $b_{\rm ph}/r_g=3 \sqrt{3}/2$)
tell us how easily one can test \ae-theory (or 
HL gravity) with gravitational lensing experiments (see in particular Ref.~\cite{GL2}
for specific attempts to use gravitational lensing to test whether astrophysical black holes are really described by GR).
Also, $b_{\rm ph}$ is related to the frequency of the circular photon orbit, $\omega_{\rm ph}=1/b_{\rm ph}$, 
which in principle will be observable with future gravitational-wave detectors because it regulates the frequency of
the black hole gravitational quasi-normal modes, at least in the eikonal approximation~\cite{eik1,eik2,eik3}.\footnote{Quasi-normal modes of black hole solutions in \ae-theory
were studied in Refs.~\cite{Konoplya:2006rv,Konoplya:2006ar} under the assumption that the aether remains unperturbed.}
\item The
ratio $r_g/r_{H}$. This quantity does not have a direct observational meaning, but 
we
compare it 
to its GR value (unity)
as a way to assess how the near-horizon region of \ae-theory (or 
HL gravity) differs from GR. This is probably not very important from an observational point of view 
(because the
near-horizon 
region 
is hard to observe), but is nevertheless conceptually interesting.
\end{enumerate}

\begin{figure}
\includegraphics[type=pdf,ext=.pdf,read=.pdf,width=8.5cm]{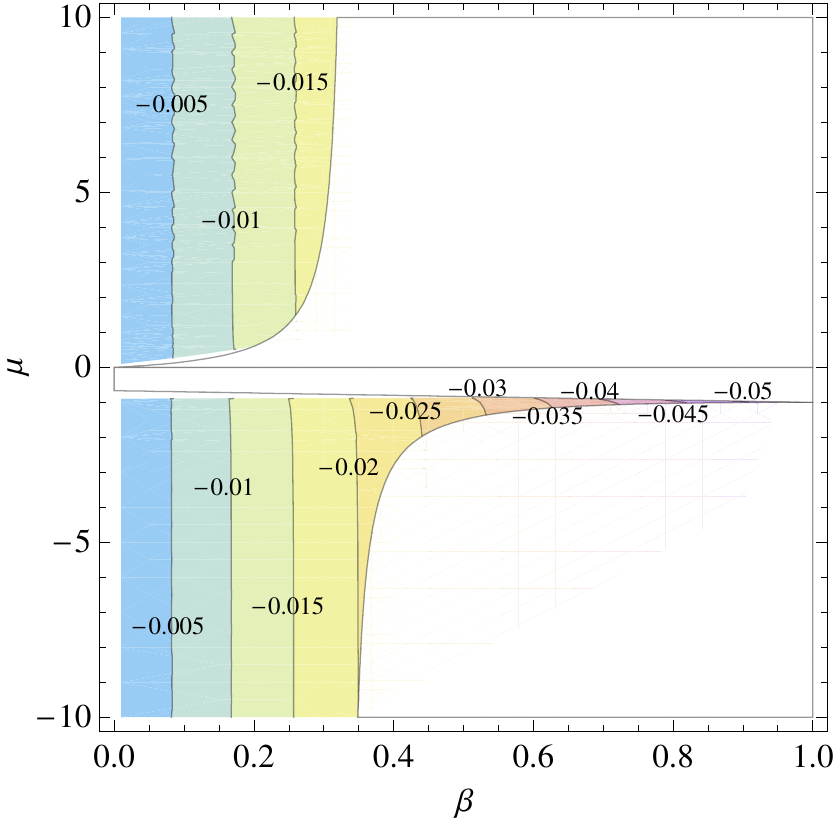}
\caption{\label{wiscoHL} Fractional  
HL gravity 
deviation from GR  for  the dimensionless product $\omega_{_{\rm ISCO}}r_g$ of the
ISCO frequency and the gravitational radius,} 
in the viable region of the parameter plane (see Sec.~\ref{parameterspace}).
 \end{figure}

 \begin{figure}
 \includegraphics[type=pdf,ext=.pdf,read=.pdf,width=8.5cm]{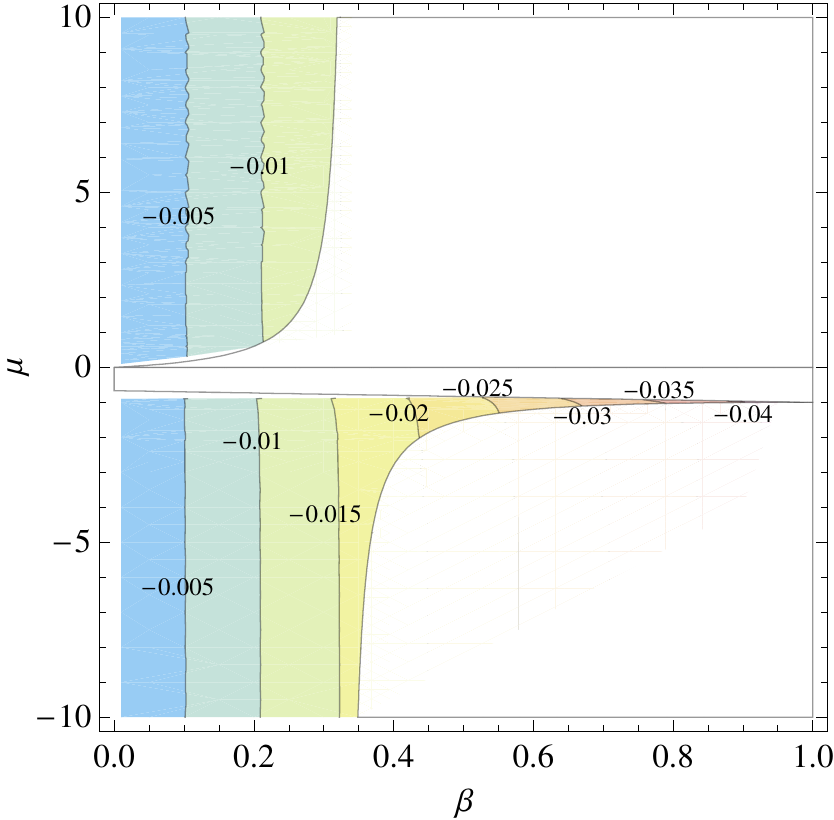}
 \caption{\label{zmaxHL} Same as in Fig.~\ref{wiscoHL}, but for 
 the maximum redshift of a photon emitted by a source moving on the ISCO, 
$z_{\rm max}=\nu_{\rm emitted}/\nu_{\rm measured}-1$, This maximum redshift is 
obtained for a photon emitted ``backward'' with respect to the velocity of the source.
 }
 \end{figure}

\begin{figure}
\includegraphics[type=pdf,ext=.pdf,read=.pdf,width=8.5cm]{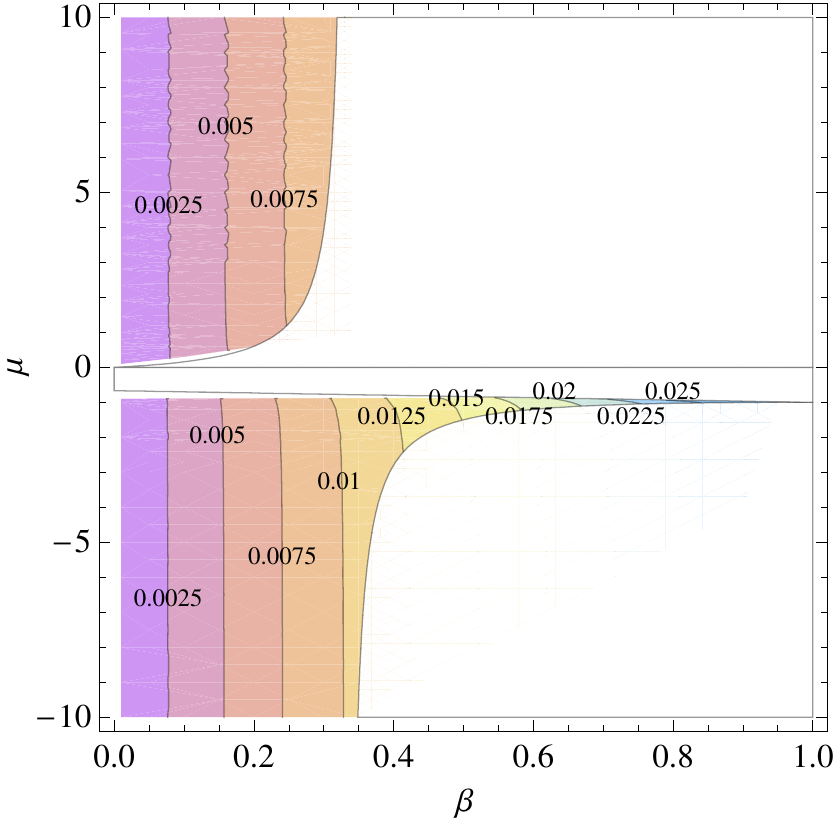}
\caption{\label{bphHL} Same as in Fig.~\ref{wiscoHL}, but for the 
impact parameter of the circular photon orbit $b_{\rm ph}$, 
normalized against  the gravitational radius $r_g$.
Note that the frequency of the circular photon orbit is $\Omega_{\rm ph}=1/b_{\rm ph}$.
}
\end{figure}

\begin{figure}
\includegraphics[type=pdf,ext=.pdf,read=.pdf,width=8.5cm]{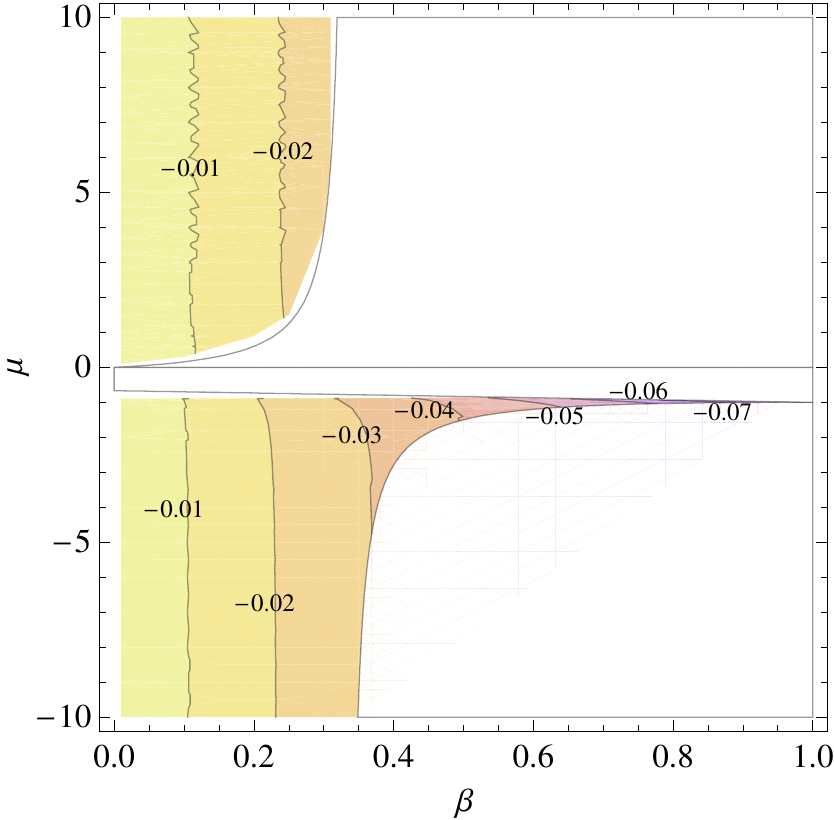}
\caption{\label{massHL} Same as in Fig.~\ref{wiscoHL}, but for the 
ratio
 $r_g/r_{H}$ of the gravitational radius $r_g$ to the horizon's circumference radius $r_H$.}
\end{figure}

To summarize our results, we have used Mathematica to produce 
contour plots for the
relative
deviations of these quantities away from their GR values, as a function
of the theory's parameters --- $(c_{+},c_{-})$ for \ae-theory and $(\beta,\mu)$ 
for HL gravity. The plots are shown in Figs.~\ref{wiscoAE} to \ref{massAE} for \ae-theory and in Figs.~\ref{wiscoHL} to \ref{massHL} for HL gravity. 
The small wiggles  
on
some of the  
contours
are simply a spurious effect of the numerical procedure used to calculate them.

Two things are striking about these plots. First, the deviations are largely
controlled by a single parameter, $c_+$ in \ae-theory and 
$\b$ in HL gravity. These two parameters are actually the same 
when the two theories are identified [cf. Eq.~(\ref{blastoae})]. 
Second, the contours for the four different
quantities in \ae-theory or HL gravity are very similar.
This could be understood if, to a good approximation, the metric function $F$
actually depended on the two coupling parameters
only via a single combination.
This combination should reduce approximately to $c_{+}$ in the \ae-theory
case and to $\beta$ in the HL gravity case
to explain the fact that the contours are almost
vertical. Using our numerical solutions, we have verified that this is
indeed the case. This may seem in contrast with the presence of a term
$c_{14}/r^3$  in the asymptotic metric \eqref{asyF}, because  the lines
$c_{14}=$constant are very different from the contours shown in Figs.
\ref{wiscoAE}-\ref{massHL}. (In the case of ae-theory, for instance, we have
$c_{14}=2c_+c_-/(c_++c_-)$ when the PPN parameters vanish, so the
$c_{14}$=constant lines  are
much more horizontal than our contours.)  However, we have verified
that while the $c_{14}/r^3$ term is indeed present in our numerical
solutions, it is always
negligible outside the horizon. This is because that term is
suppressed by a factor $1/48$ [\textit{cf.} Eq.  \eqref{asyF}]
and because  $\vert c_{14} \vert \lesssim 2$ in the part of parameter
space that we explore, both in ae-theory and HL gravity.
Higher order terms in the $1/r$ expansion are, however, important,
and are responsible for producing the deviations from GR in
the dimensionless quantities plotted in the figures (the deviation of
$r_g/r_H$ from its GR value is also sensitive to the $1/r$ term).

As can be seen, the deviations from GR are quite small. In \ae-theory, they do not reach the 10\% level until
$c_+\approx0.7$, while in HL gravity they remain less than
3\% in most of the parameter space.
Such small deviations
are probably undetectable with accretion disk 
spectra, iron-K$\alpha$ lines, or gravitational lensing, at least with present data. 
% anytime in the 
% near future. 
The only exception might be the 
region $0.9<c_{+}<1$ in the \ae-theory
parameter plane, where the deviations from GR are larger than $20$\% 
and might therefore be observable with these techniques,
provided that systematic errors in the data and astrophysical uncertainties are properly understood.
However, this region is presumably already ruled out by observations of gravitational radiation
damping from binary pulsars~\cite{Foster:2007gr}. 

The size of the black hole modifications we have found is comparable to the 
effects on neutron star structure reported in Ref.~\cite{Eling:2007xh}. 
(The solution outside a star, in which the aether is aligned with the 
timelike Killing vector, is different from the black hole solution,
in which the aether flows inward.) 
There the maximum 
mass, surface redshift, and ISCO frequency were computed for various equations of state.
Depending on the equation of state, the maximum masses
are about 6\% - 15\% smaller than in GR when the
\ae-theory parameter $c_{14}$
is equal to 1, and the corresponding
surface redshifts are roughly 10\% larger than in GR. The ISCO frequency, which is
independent of the equation of state, is only 4\% smaller than in GR. 
Thus, it also appears challenging to obtain useful constraints from 
neutron star observations.

On the other hand, future gravitational-wave experiments such as LISA will be able test deviations from
the Kerr metric with astonishing accuracy ($\sim10^{-6}$) 
using extreme mass-ratio inspirals (EMRIs)~\cite{emris_review}. 
While further work is necessary 
to be ready to use LISA data for testing 
\ae-theory or 
HL gravity, such an accuracy is more than enough (by orders of magnitude) to detect 
 the deviations of \ae-theory (HL gravity) from GR predicted by our code, essentially in all of the parameter space. 
Observations of 
EMRIs with LISA could, therefore, allow very strong constraints to be put on \ae-theory and 
HL gravity.

\subsection{The interior solutions}

In LV theories different modes generically propagate at different velocities. Additionally, they no longer necessarily satisfy linear dispersion relations, which means that  the limiting speed of short wavelength disturbances can be infinite in the preferred frame. Indeed \ae-theory has the first of these 
characteristics and (full) HL gravity has both of them, as we will discuss in more detail below. In such theories a mode can potentially escape the horizon defined by another mode, or there might not be any horizons at all. 
 It is therefore potentially of observational 
interest to study the interior behavior of the black hole
solutions. It is also in any case of mathematical and fundamental interest.
We 
present here some results for this behavior.

\begin{figure}[t]
\includegraphics[type=pdf,ext=.pdf,read=.pdf,width=8.5cm]{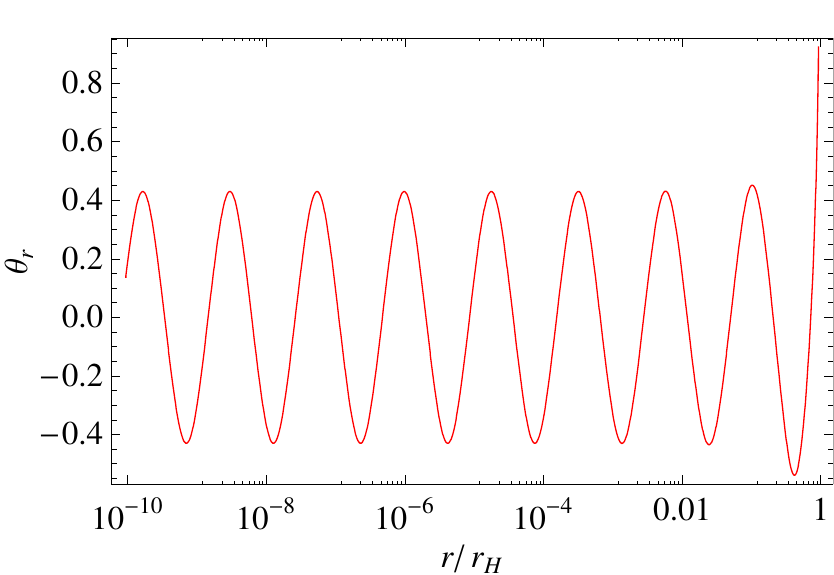}
\caption{\label{theta_r_inner1} The aether's boost angle $\theta_r= {\rm arccosh}\gamma_r$ with
respect to the future-directed observer orthogonal to the hypersurfaces $r=$ const,
as a function of radius and inside the
metric horizon ($r=r_H$). This is for \ae-theory with $c_-=0.0018$, $c_+=0.01$. 
}
\end{figure}

\begin{figure}[t]
\includegraphics[type=pdf,ext=.pdf,read=.pdf,width=9.5cm]{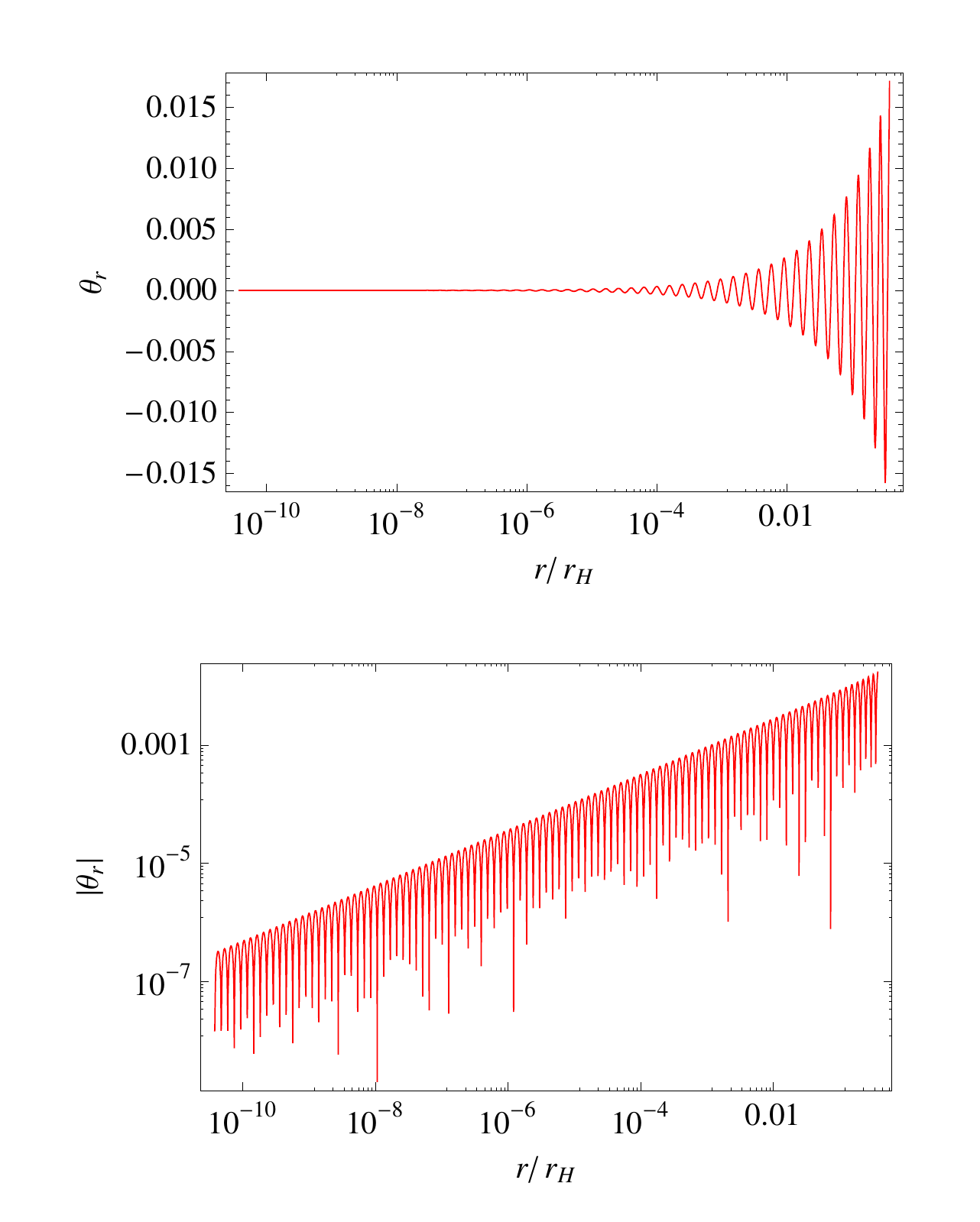}
\caption{\label{theta_r_inner2} The same as in Fig.~\ref{theta_r_inner1}, but for \ae-theory with $c_+=0.99$, $c_-=0.01$. 
}
\end{figure}

\begin{figure}[t]
\includegraphics[type=pdf,ext=.pdf,read=.pdf,width=9.5cm]{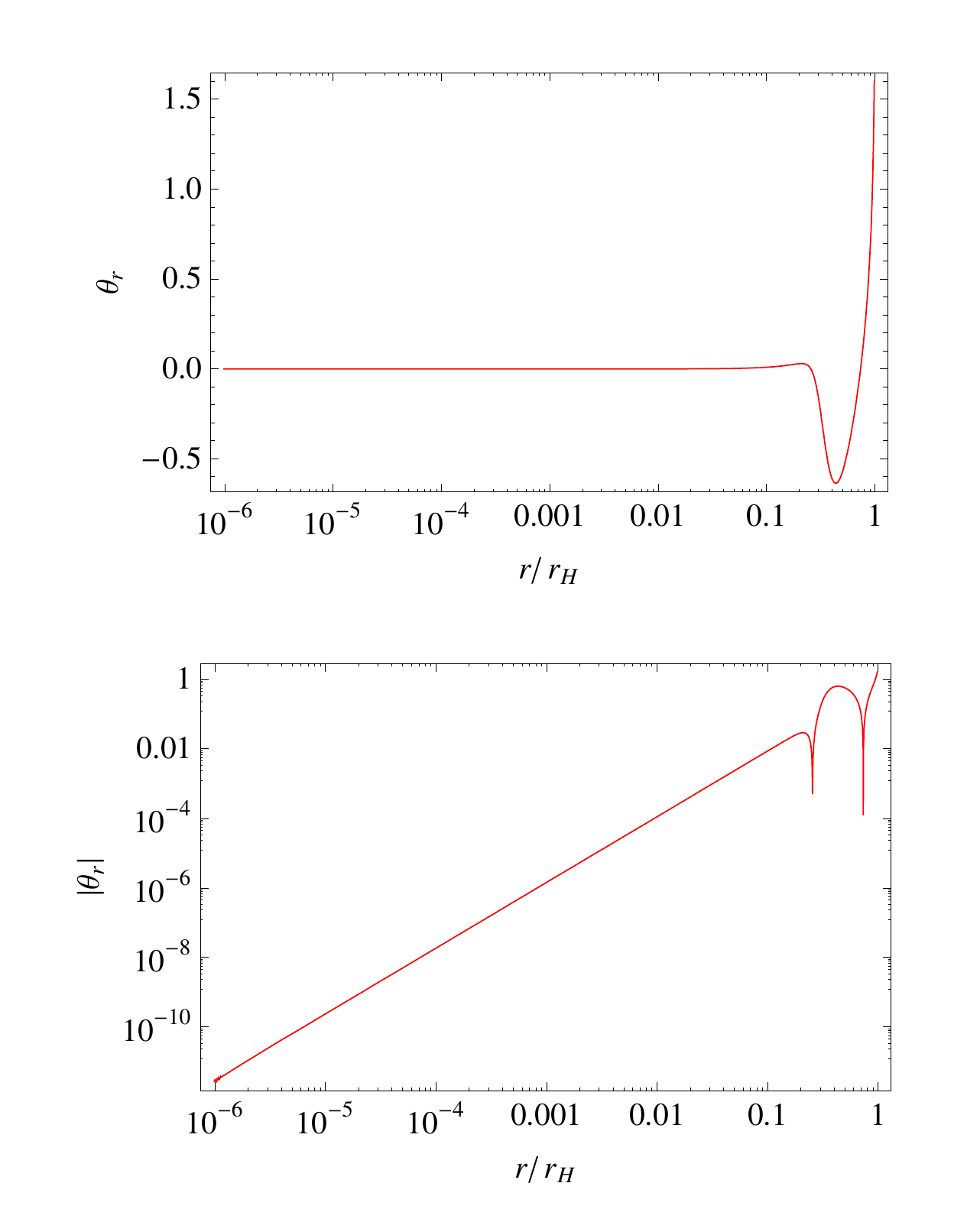}
\caption{\label{theta_r_inner3} The same as in Fig.~\ref{theta_r_inner1}, but for HL gravity with $\beta=0.4$, $\mu=-1.8$. 
}
\end{figure}

In general, irrespective of the parameters $(c_{+},c_{-})$ or $(\beta,\mu)$ of the theory, 
the solutions that we have found always present
a spacelike
curvature singularity at $r=0$, where the Ricci scalar $R$ and the Kretschmann scalar
$R_{\alpha\beta\mu\nu}R^{\alpha\beta\mu\nu}$ (as well as the metric function $F$) diverge. 
The aether, to the contrary,
presents an oscillatory behavior, as already noticed in Ref.~\cite{Eling:2006ec}. 
To see this, we have studied the Lorentz factor 
$\gamma_r\equiv{u}^\alpha_{\rm obs}u_\alpha$
of the aether as measured by the future directed observer orthogonal to the 
(spacelike) hypersurface $r=$ constant.
In Figs. \ref{theta_r_inner1}-\ref{theta_r_inner3}
we plot the corresponding boost angle 
\beq\label{boostangle}
\theta_r={\rm arccosh} \gamma_r,
\eeq
for three representative cases,\footnote{These cases seem qualitatively representative of the inner solutions
of all of the regular black holes that we studied, although we 
have not yet performed
a full 
parameter space 
scan like those 
presented in the previous section.} 
as a function of $r$ (which plays the role of a time coordinate inside the horizon).

For the case in Fig.~\ref{theta_r_inner1} (\ae-theory with $c_-=0.0018$, $c_+=0.01$),  
after an initial transient, the boost angle oscillates with roughly constant amplitude, 
and roughly constant period when measured in terms of 
$\log r$,  
thus corresponding to an undamped oscillator. 
The case shown in Fig. \ref{theta_r_inner2} (\ae-theory with $c_+=0.99$, $c_-=0.01$)
resembles instead a damped oscillator, while the case shown in  Fig. \ref{theta_r_inner3} (HL gravity with $\beta=0.4$, $\mu=-1.8$) 
resembles, after an initial transient, an over-damped 
oscillator (\textit{i.e.} one in which the damping 
is so strong that it does not oscillate, but approaches 
the rest position exponentially). 

This oscillatory behavior is reminiscent of that found for the aether in 
(anisotropic)
Bianchi type I (Kasner-like) cosmologies with a cosmological constant,
both in HL and \ae-theory~\cite{Carruthers:2010ii}. 
This analogy is understandable, since the black hole interior also corresponds to
an anisotropic cosmology, although of a different symmetry type and spatial curvature,
and without a cosmological constant. As remarked in Ref.~\cite{Eling:2006ec},
it is also reminiscent of oscillations found in the interior of Einstein-Yang-Mills
black hole solutions~\cite{Donets:1996ja}. 
Presumably the appearance of oscillations can be 
understood from the cosmological point of view as an attractor arising because
of a ``restoring force'' that tends to align the cosmological rest frame with any
other preferred frame defined by the other (non-metric) fields.

There is an important causal implication of these oscillations, namely, 
they imply that the concept of 
a
black hole survives in these
theories, as we now explain \cite{Sergey}.  
In \ae-theory, the spin-0, spin-1, and spin-2 perturbations
generally all travel at different speeds, and we have assumed they are all
greater than or equal to the metric speed of light in order to satisfy the
vacuum \v Cerenkov constraints. Such perturbations could therefore 
escape from a metric horizon, although  they might have deeper causal horizons
trapping them. 
 The same will hold for the spin-0 and spin-2 perturbations in the low energy limit of HL gravity. 
However, the situation is quite different when the full HL theory is considered, {\em i.e.}~when the higher 
order terms in $L_4$ and $L_6$ (and presumably the corresponding higher order spatial 
derivative terms for matter fields)  are also taken into account.  Because these terms contain higher spatial derivatives,
short wavelength perturbations can travel at arbitrarily high speed relative to the
aether. This raises the possibility that in HL gravity signals can always escape from 
{\it anywhere} inside a black hole, in which case the concept of black hole 
really would not survive at all in HL gravity.

However,
even at infinite speed, signals cannot travel backward in time,
{\em i.e.}~backward across the  
constant $T$ hypersurfaces which are orthogonal to the
aether (\ref{ho}). These hypersurfaces define a universal causal structure in 
HL gravity, and in hypersurface-orthogonal aether configurations.
When the boost angle (\ref{boostangle}) vanishes, 
the aether is orthogonal 
to a constant-$r$ hypersurface $r=r_c$, which
therefore coincides with a constant-$T$ hypersurface. 
Events at $r<r_c$ (which lie to the future of $r_c$) 
therefore can have no influence on 
events at $r>r_c$, so signals are trapped within 
the radius $r_c$ \cite{Sergey}.
In fact the oscillatory behavior means that there
are many such surfaces
inside the black hole. In the 
examples shown in Figs.~\ref{theta_r_inner1}-\ref{theta_r_inner3}, 
the first such surface
occurs quite close to the metric horizon, as measured by $r$. 
We refer to the outermost such surface as the ``universal horizon".

The HL time function $T$ therefore has a peculiar behavior in the
black hole interior. A surface of constant $T$ that comes in from
spatial infinity crosses the metric horizon
smoothly, but then, unlike the familiar Painlev\'e-Gullstrand time coordinate,
rather than running into the singularity at $r=0$ this surface dips down to
the infinite past, measured in the advanced time coordinate $v$, at the
universal horizon.
This behavior of $T$
happens 
despite the fact that the aether vector and the geometry
are perfectly smooth all the way up to the singularity.
To understand what is happening,
we note that 
spherical  and time translation symmetry imply that $T$ has
the form $T(v,r)=h(v+f(r))$, where $f(r)$ is determined by the
metric and aether, and
the function
$h$ is completely arbitrary (other than
that it be monotonic) on account of
the $T$-reparametrization symmetry 
of the theory, Eq. (\ref{fpd}). It turns out that
the function $f(r)$ diverges at the universal horizon (and at the
similar surfaces inside it).
Thus, on a constant $T$ surface, $v\to -\infty$
as the universal horizon is approached. Also, on an ingoing light ray
at constant $v$, the function $T$ diverges unless $h$ is chosen
properly. For instance, we could choose  $h$ to be  minus the inverse of $f$,
in which case  $T$ remains finite and increasing 
when approaching the universal horizon from the outside
at constant $v$.

A final important comment is that the solutions we have been discussing 
apply only in the IR limit of HL gravity, \textit{i.e.}~they neglect the effects of 
higher derivative terms in the HL action 
(\ref{SBPSHfull}). 
Those terms will certainly strongly 
affect the solution in the black hole interior where the curvature becomes 
large. However, as remarked above, the universal horizon appears 
fairly close to the metric horizon, in a region where one would expect 
the IR limit of the theory to still be an excellent approximation for 
macroscopic black holes. Hence, there is good reason to 
expect 
that sufficiently
large black holes in the full HL theory also possess a universal horizon. 

\section{Conclusions}

\label{concs}

We have studied static, spherically symmetric, black-hole solutions in Einstein-aether theory. These are also solutions of the low-energy limit of Ho\v rava--Lifshitz gravity. We highlight below the most 
important steps in the process of finding these solutions (setting aside technical issues) 
and summarize their salient properties.

In general these spacetimes have
both a
metric horizon and a horizon for the spin-0 mode of the aether, as the latter travels at different speed than the speed of light defined by the metric cone. 
The vacuum \v Cerenkov constraint requires that the speed of the spin-0 mode be greater than
or equal to
 the speed of light, so we only considered cases where the spin-0 horizon lies inside the metric horizon. Additionally, we have imposed the condition that 
 the spin-0 
 horizon be regular, as this is what is expected for black holes that form from gravitational collapse.

Imposing this regularity condition and the condition of asymptotic flatness leads to a one parameter family of solutions for each set of the parameters of the theory. 
In units where the horizon radius is $1$ this yields
a unique solution for that set of theory parameters. We have generated this solution numerically. A crucial step towards this has been the realization that a specific combination of the field equations constitutes  a set of constraint equations. Additionally, a suitable field redefinition has 
been utilized in order to facilitate imposition of the condition of regularity at the spin-0 horizon.

We have restricted 
attention to the regions of the parameter space, both in Einstein-aether theory 
and Ho\v rava--Lifshitz gravity, where the following conditions are satisfied: (i) stability and positive energy of all perturbations; (ii) avoidance of vacuum \v Cerenkov radiation; (iii) exact agreement with 
 the first post-Newtonian predictions of GR
(vanishing preferred frame parameters). 
(The
regions of the parameter space where these 
conditions are satisfied do not coincide for the two theories.)

We find that regular, asymptotically flat black-hole solutions exist
in all of the  
explored parameter regions
satisfying these observational constraints.
In order to characterize the solutions, we used 
 three dimensionless quantities that are in principle measurable by observations: the 
product of the orbital frequency of the Innermost Stable Circular Orbit (ISCO) with the gravitational radius, 
the maximum redshift of a photon emitted by a source moving on the ISCO, and the  
ratio between the impact parameter of the circular photon orbit and the gravitational radius. 
As a fourth probe we used 
the ratio between the gravitational radius and the horizon radius. 

The deviations of these quantities from the values that they have in general relativity are always quite small, exceeding 10\% (in Einstein-aether theory) and 3\% (in Ho\v rava--Lifshitz gravity) only in restricted portions of the parameter space. Therefore, they are expected to be difficult, 
although perhaps not impossible, to detect by electromagnetic observations such as accretion disk spectra, iron-K$\alpha$ lines or gravitational lensing. However, future gravitational-wave experiments, such as LISA, would have more than sufficient accuracy to detect these deviations from general relativity. 

Last but not least, we considered the interior solution. 
Inside the black hole the aether oscillates with respect to the constant $r$ surfaces, and there is a spacelike
singularity at $r=0$. More importantly, 
it turns out that there exists 
 a universal horizon
inside the metric and spin-0 horizons, {\em i.e.}~a 
surface of constant $r$ that is orthogonal to the aether.
No modes can escape from inside the universal horizon, 
even those 
satisfying modified dispersion relations that could allow them to travel 
at arbitrarily high speeds relative to the aether, as expected in Ho\v rava--Lifshitz gravity. 
The existence of this universal horizon implies that the concept of 
a
black hole, 
in the sense of a region of spacetime where all signals are trapped, 
seems to survive in these theories.

\acknowledgments
E.B. acknowledges support from NSF Grant PHY-0903631. 
TJ was supported in part by the NSF under grant PHY-0903572. TPS was supported by a 
Marie Curie Fellowship. \\
\appendix*
\def\rs{r_{_{\rm ISCO}}}

\section{Quantities used to characterize the solutions}

The innermost stable circular orbit (ISCO) occurs 
at the radius $r=\rs$ where 
\beq
-2 r F'^2 + F(3 F' + r  F'')=0.
\eeq
The ISCO frequency is 
\beq
\o_{_{\rm ISCO}}=\left.\sqrt{\frac{F'}{2 r}}\right\vert_{_{\rm ISCO}}.
\eeq

The redshift $z=\nu_0/\nu_\infty-1$
of a photon emitted by a source orbiting
at the ISCO and measured at infinity
is given by 
\beq  
1+ z =  \left.\frac{1 - \omega b}{\sqrt{F -\omega^2 r^2}}\right\vert_{_{\rm ISCO}},
\eeq
where $b$ is the impact parameter of the photon, which is
defined as its angular momentum divided by its energy,
and which characterizes the direction of emission.
For instance, for a photon transverse to the orbit one has $b=0$,
while for photons emitted in the forward or backward directions one has $b=\pm (r/\sqrt{F})_{\rm ISCO}$.
The maximum redshift $z_{\max}$ is therefore in the backward direction, 
\beq  
1+ z_{\max} =   \left.\frac{1 + \omega rF^{-1/2}}{\sqrt{F -\omega^2 r^2}}\right\vert_{_{\rm ISCO}},
\eeq
and this is because the Doppler shift is maximized in this case.

The circular photon orbit occurs at $r=r_{\rm ph}$, where
\beq
-2 F + r F'=0.
\eeq
The impact parameter of the photon orbit is given by
\beq
b_{\rm ph} = \left.\frac{r}{\sqrt{F}}\right\vert_{\rm ph},
\eeq
while its frequency is $\omega_{\rm ph}=1/b_{\rm ph}$.

The relative Lorentz gamma factor between the 
aether and the unit Killing energy observer at the horizon,
which corresponds to an observer that falls in radially
from rest at infinity,  
is
\beq
\gamma_{\rm ff}=u_\mu u_{\rm ff}^\mu = A_H +\frac{1}{4A_H},
\eeq
when the $v$ coordinate is normalized so that $F_\infty=1$.
The relative gamma factor between the aether and the 
unit (timelike) normal to the constant $r$ surfaces inside the 
metric horizon is 
\beq
\gamma_r=- \frac{u^r}{\sqrt{g^{rr}}}
\eeq
from which we define the boost angle as $\theta_r={\rm arccosh}(\gamma_r)$.

\end{document}